\documentclass[lettersize,journal]{IEEEtran}
\usepackage{amsmath,amssymb,amsfonts}
\usepackage{cite}
\usepackage{algorithmic}
\usepackage{array}
\usepackage{textcomp}
\usepackage{float}
\usepackage{stfloats}
\usepackage{url}
\usepackage{verbatim}
\usepackage{graphicx}
\usepackage{subfigure}
\usepackage{algorithm}
\usepackage[export]{adjustbox}
\def\BibTeX{{\rm B\kern-.05em{\sc i\kern-.025em b}\kern-.08em
    T\kern-.1667em\lower.7ex\hbox{E}\kern-.125emX}}
\usepackage{balance}

\newtheorem{lemma}{\bf Lemma}

\begin{document}
\bibliographystyle{IEEEtran}
\title{Joint Optimization of Communication Enhancement and Location Privacy Protection in RIS-Assisted Underwater Communication System}
\author{Ziqi Chen, \IEEEmembership{Graduate Student Member, IEEE}, Jun Du, \IEEEmembership{Senior Member, IEEE}, Chunxiao Jiang, \IEEEmembership{Fellow, IEEE} and Zhu Han, \IEEEmembership{Fellow, IEEE}\vspace{-8mm}
	\thanks{Z. Chen and J. Du are with the Department of Electronic Engineering, Tsinghua University, Beijing 100084, China (e-mail: 18801351390@163.com, jundu@tsinghua.edu.cn.}
	\thanks{Chunxiao Jiang is with the Beijing National Research Center for Information Science and Technology and the Tsinghua Space Center, Tsinghua University, Beijing 100084, China. (email: jchx@tsinghua.edu.cn).}
	\thanks{Z. Han is with the Department of Electrical and Computer Engineering in the University of Houston, Houston, TX 77004 USA, and also with the Department of Computer Science and Engineering, Kyung Hee University, Seoul. South Korea, 446-701 (e-mail: hanzhu22@gmail.com).}
}

\markboth{Journal of \LaTeX\ Class Files,~Vol.~18, No.~9, September~2020}%
{How to Use the IEEEtran \LaTeX \ Templates}

\maketitle

\begin{abstract}
As the demand for underwater communication continues to grow, underwater acoustic RIS (UARIS), as an emerging paradigm in underwater acoustic communication (UAC), can significantly improve the communication rate of underwater acoustic systems. However, in open underwater environments, the location of the source node is highly susceptible to being obtained by eavesdropping nodes through correlation analysis, leading to the issue of location privacy in underwater communication systems, which has been overlooked by many studies. To enhance underwater communication and protect location privacy, we propose a novel UARIS architecture integrated with an artificial noise (AN) module. This architecture not only improves communication quality but also introduces noise to interfere with the eavesdroppers' attempts to locate the source node. We derive the Cram\'er-Rao Lower Bound (CRLB) for the localization method deployed by the eavesdroppers and, based on this, model the UARIS-assisted communication scenario as a multi-objective optimization problem. This problem optimizes transmission beamforming, reflective precoding, and noise factors to maximize communication performance and location privacy protection. To efficiently solve this non-convex optimization problem, we develop an iterative algorithm based on fractional programming. Simulation results validate that the proposed system significantly enhances data transmission rates while effectively maintaining the location privacy of the source node in UAC systems.
\end{abstract}


\section{Introduction}
In recent years, with the flourishing development of the marine economy, underwater Internet of Things devices based on acoustic communication, such as underwater sensors and Autonomous Underwater Vehicles (AUVs), have become extremely important for the exploration and utilization of the ocean. Meanwhile, underwater acoustic communication (UAC) has a wide range of applications, including real-time data collection on water, natural disaster prevention, marine resource exploration, and maritime security \cite{mohsan2023recent}, \cite{ghazy2024irs}. However, compared to terrestrial wireless communication systems, the data rate of UAC systems is so low, remaining at the kilobits per second (kbps) level, that it is difficult for UAC to support the normal operation of most underwater applications \cite{zhu2023internet}, \cite{cao2020channel}. This low data rate issue is caused by the characteristics of the underwater environment, particularly in shallow water, where the low bandwidth of sound waves and severe multipath fading contribute to the problem \cite{feng2021message}.

To ensure the reliability and stability of UAC systems, many existing studies aim to improve the transmission rate of these systems through underwater acoustic reconfigurable intelligent surfaces (UARIS). Specifically, UARIS can be deployed beneath ships, buoys, or AUVs for underwater operations \cite{dong2023reconfigurable}. The UARIS is capable of altering the amplitude and phase of incident acoustic signals and generating strongly reflected paths through beamforming \cite{bo2024dual}. Compared to natural UAC channels, this newly formed reflected path can significantly enhance the signal-to-noise ratio (SNR) at the receiver, thereby achieving higher data transmission rates.

Many existing studies focus on enhancing communication in underwater environments, but few have considered the issue of location privacy in underwater communication. Due to the openness of the underwater communication environment, the source node (SN) of the transmitted signal is highly susceptible to being located by illegal eavesdropping nodes (ENs) through analysis of the received signals, thereby threatening the privacy and security of the node. In this context, the challenge lies in leveraging Reconfigurable Intelligent Surface (RIS) technology to enhance underwater communication while protecting the location information of the source. Some studies achieve interference with ENs through the synergy of artificial noise (AN) from the transmitter and RIS \cite{han2022artificial}, \cite{arzykulov2023artificial}. However, these studies assume that the source and RIS have prior knowledge of the eavesdropper location, which does not align with real-world eavesdropping scenarios.

Therefore, we add an AN module to the UARIS, enabling the UARIS to intelligently control the reflected signals while ensuring communication quality, and simultaneously introduce additional noise signals to interfere with the attempts of ENs to infer the SN location. The main contributions of this paper are summarized as follows:

The main contributions of this work are outlined as follows:
\begin{itemize}
	\item We propose a UARIS architecture with an AN module. This module introduces noise into the reflected signals to interfere with illegal ENs, preventing them from locating the SN. This design overcomes the limitation of previous assumptions that required prior knowledge of the locations of the ENs, enhancing privacy protection in real-world eavesdropping scenarios.
	\item We design a mechanism for underwater communication enhancement and location privacy protection based on acoustic RIS and AN. Specifically, by intelligently controlling the amplitude and phase of the reflected signals, UARIS not only improves the SNR at the receiver, enhancing communication performance but also introduces dynamically controllable AN to protect the location privacy of the SN. This mechanism not only enhances communication stability but also addresses the need to protect the location privacy of the SN.
	\item We model underwater communication enhancement and location privacy protection as a multi-objective optimization problem. This problem optimizes transmission beamforming, reflective precoding, and noise factors to maximize communication performance while protecting the privacy of the SN. The paper provides a detailed derivation of the Signal-to-Interference-plus-Noise Ratio (SINR) calculation and analyzes the impact of the noise factor on location privacy based on the Cram\'er-Rao Lower Bound (CRLB).
	\item To address the non-convexity of the optimization problem, this work employs fractional programming (FP) to decouple it and proposes an iterative algorithm for jointly optimizing the transmission beamforming, reflection matrix, and noise factor. In each iteration, the auxiliary variables, noise factor, beamforming vector, and reflection matrix are optimized step by step, gradually converging to a locally optimal solution, thereby improving the algorithm's efficiency and stability.
\end{itemize}

The structure of this paper is as follows: Section \ref{work} introduces the related works. Section \ref{sys} presents a detailed exposition of a downlink shallow water communication scenario based on UARIS. Section \ref{SBL} describes a semi-blind localization method for ENs without prior channel information and conducts a CRLB analysis. Section \ref{opt} designs an optimization scheme for joint transmit beamforming, reflective precoding, and noise factor. Section \ref{sim} validates the performance and relevant metrics of the proposed mechanism. Section \ref{concl} is the conclusion of the entire paper. 

\emph{Notations:} $\mathbb{C}$, $\mathbb{R}$ and $\mathbb{R}_+$ denote the sets of complex, real and positive real numbers, respectively. $\left[\cdot\right]^\top$, $\left[\cdot\right]^*$, $\left[\cdot\right]^\dagger$ and $\left[\cdot\right]^{-1}$ represent the transpose, conjugate, conjugate-transpose and inverse operations of a matrix, respectively. $\|\cdot\|$ and $\|\cdot\|_F$ denote the Euclidean norm and the Frobenius norm of the argument, respectively. $\mathbf{I}_K$ is the $K\times K$ identity matrix. $\mathbf{e}_k$ denotes the $k$-th column of the identity matrix. $\mathcal{R}(\cdot)$ and $\mathcal{I}(\cdot)$ denote the real and imaginary parts of the complex-valued arguments, respectively. $\lambda_{\max}(\cdot)$ represents the largest eigenvalue of the matrix. $\text{Diag}(\cdot)$ forms an $K\times K$ diagonal matrix from a $K$-dimensional vector argument. $\mathcal{CN}\left(\mathbf{\mu}, \mathbf{\Sigma}\right)$ denotes the complex multivariate Gaussian distribution with mean $\mathbf{\mu}$ and variance $\mathbf{\Sigma}$.

\section{Related Works}
\label{work}
\subsection{Reconfigurable Intelligent Surface}
RIS is an emerging paradigm in the field of wireless communications that controls and optimizes signal propagation paths by intelligently adjusting the phase and amplitude of its reflective elements \cite{elmossallamy2020reconfigurable}. Comprising a large number of controllable reflective units, RIS can dynamically alter the direction, strength, and coverage of electromagnetic waves, thereby enhancing signal quality, improving communication efficiency, and reducing interference. Specifically, Huang $et~al.$ \cite{huang2019reconfigurable} proposed a multi-user MISO system based on passive RIS, significantly improving the energy efficiency by optimizing the power allocation of the base station and the passive RIS reflection phase design. However, during the research on passive RIS, it was found that RIS introduces a "multiplicative fading effect," where the path loss of the transmitter-RIS-receiver link is the product of the path losses of the transmitter-RIS link and the RIS-receiver link (rather than their sum), resulting in the gain of the reflected link being much smaller than that of the direct link \cite{liu2021active}. To overcome the multiplicative fading introduced by RIS, Zhang $et~al.$ \cite{zhang2022active} proposed an active RIS architecture integrated with reflection-type amplifiers and optimizes joint beamforming and reflective precoding in a multi-user MISO system, significantly enhancing the capacity gain of the communication system.

Different from wireless channels, the highly complex characteristics of underwater acoustic channels make the application of RIS in UAC systems more challenging. Wang $et~al.$ \cite{wang2023designing} proposed an acoustic RIS design, which enhances the capacity and reliability of underwater communication systems through new hardware, ultra-wideband beamforming, and practical operational protocols. Simulations verify that this system significantly improves data transmission rates in underwater environments, addressing the low data rate and unreliable channel issues in traditional UAC systems. However, the current designs and optimizations of UARIS are only focused on communication enhancement. To further protect the location privacy, we need to improve the existing UARIS architecture.
\subsection{Underwater Acoustic Localization Method}
Due to the complexity and high dynamics of the underwater environment, underwater acoustic localization (UAL) is highly challenging. Currently, most existing UAL methods can be divided into two categories: Direction of Arrival (DoA)-based methods and Time of Arrival (ToA)-based methods. The DoA methods estimate the position of the SN by measuring the direction of arrival of sound waves from the source to the receiver and using the angle information from multiple receivers. Zhang $et~al.$ \cite{zhang2022covariance} proposed an array aperture extension method based on covariance matrix reconstruction, which increases virtual array elements by constraining the amplitude and phase, thereby improving the precision and resolution capability of DoA estimation in underwater acoustic systems. Liu $et~al.$ \cite{liu2023unified} proposed an iterative technique for joint estimation of position and sound propagation speed in underwater environments characterized by an isogradient sound speed profile. Through stratification compensation and iterative refinement, the accuracy of the source DoA estimation was improved.

Meanwhile, the ToA method estimates the position of the SN by measuring the propagation time of sound waves from the source to the receiver and using known sound speed and measurements from multiple receiving nodes (RNs). Sun $et~al.$ \cite{sun2020underwater} proposed an underwater acoustic localization algorithm based on the Generalized Second-Order ToA to locate a black box sunk on the seabed, addressing the impact of signal period drift on localization accuracy. Li $et~al.$ \cite{li2023long} proposed a long baseline underwater acoustic localization method based on particle filtering and "track-before-detect" technology, aiming to address the direct sound selection problem in traditional ToA localization methods.

However, in most current research on UAL, the nodes performing localization often have access to a significant amount of prior knowledge, which may be challenging to obtain in practical underwater localization scenarios. Therefore, in UAC systems, an EN requires a method for localization that does not rely on extensive prior knowledge.

\subsection{Underwater Location Privacy Protection}
Whether in terrestrial wireless communication or UAC, the openness of the communication environment can expose the SN to localization attacks by unauthorized ENs. Therefore, to protect the location privacy of the SN, many studies have proposed interference and defense strategies against ENs. Fan $et~al.$ \cite{fan2024secure} proposed a secure localization scheme for underwater acoustic sensor networks based on autonomous underwater vehicle formation and cooperative beamforming to defend against eavesdropping attacks. Liu $et~al.$ \cite{liu2023uav}employed an unmanned aerial vehicle (UAV) to jam eavesdroppers by optimizing its trajectory and jamming power. It improved physical layer security by emitting AN, which degrades the eavesdropping channel while maintaining good communication with legitimate receivers. Hou $et~al.$ \cite{hou2024optimal} defended against eavesdropping localization attacks by using UAVs to emit AN and optimizing the UAV trajectory and jamming power.  

However, most studies assume that the SN can obtain the location information or even the channel information of the ENs, which is difficult to achieve in real eavesdropping scenarios. Therefore, we need a method that can perform jamming without requiring any prior knowledge of the ENs to protect the location privacy of the SN.

\section{System Model}
\label{sys}
\begin{figure}[!t]
	\centering
	\includegraphics[width=3.4in]{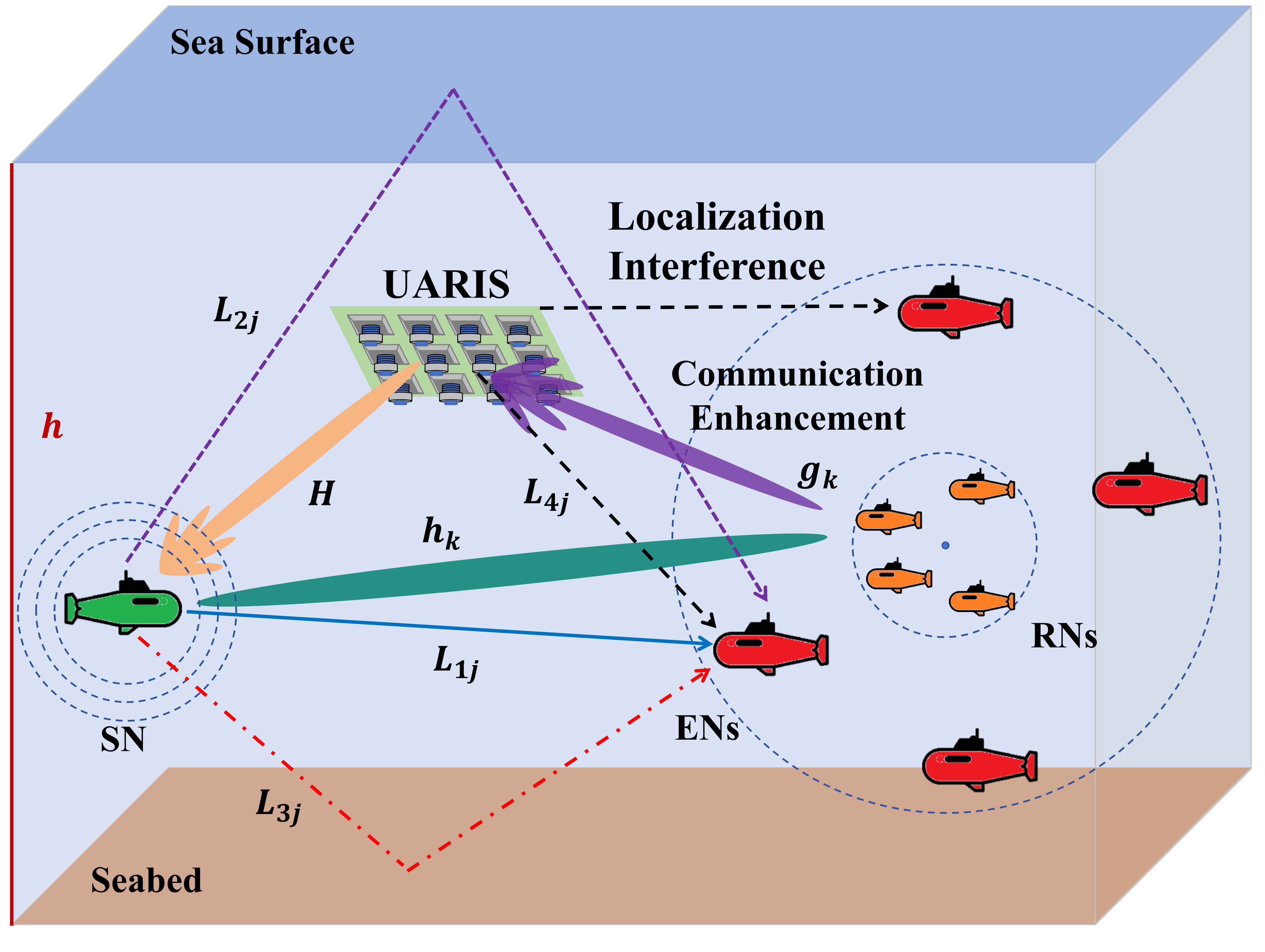}
	\caption{The downlink UARIS-assisted shallow water communication scenario.}
	\label{FRAMEWORK}
\end{figure}

In this paper, we consider a typical downlink UARIS-assisted shallow water communication scenario, which is shown in Fig. \ref{FRAMEWORK}, consisting of one SN with $T$ underwater antennas, one UARIS with $M$ reflection elements, and $K$ legitimate single-antenna RNs, of which the set is $\mathcal{K} = \left\{1,2,\cdots,K\right\}$. The position coordinates of these three entities are defined as $\mathbf{p}_S\triangleq\left[x_S\ y_S \ z_S\right]^\top$, $\mathbf{p}_{U}\triangleq\left[x_{U}\ y_{U} \ z_{U}\right]^\top$ and $\left\{\mathbf{p}_{R,k} \triangleq\left[x_{R,k}\ y_{R,k} \ z_{R,k}\right]^\top, k\in\mathcal{K}\right\}$, respectively, and assume that the three can share location information. Additionally, in the open shallow water environment, there are $J$ ENs that attempt to illegally locate the SN by analyzing the received signals. The set of ENs is denoted as $\mathcal{J}=1,2,\cdots,J$, and their position are defined as $\left\{\mathbf{p}_j\triangleq\left[x_j\ y_j \ z_j\right]^\top, j \in\mathcal{J}\right\}$. We assume that the SN is static and located sufficiently far from the RNs and all ENs, allowing for a plane wavefront (far-field) approximation in the shallow water waveguide.

In this communication scenario, we assume ENs can share all information, including the received signals and their respective location. Furthermore, the ENs can obtain the location of the UARIS, while the UARIS cannot obtain the location information of the ENs. 
\subsection{Underwater Acoustic RIS}
The UARIS enhances underwater communication by absorbing and radiating sound waves, with its working mechanism summarized as "absorption first, radiation later." Specifically, each reflection unit of the UARIS consists of two piezoelectric elements. When sound waves impinge on the surface of the acoustic UARIS, the piezoelectric elements first absorb the acoustic energy and adjust the phase and amplitude of the reflected sound waves through internal circuits. The adjusted sound waves are then radiated by another piezoelectric element, thereby achieving intelligent control of the sound waves. Additionally, each reflection unit is integrated with a reflection-type power amplifier, allowing the UARIS to amplify the signal while reflecting it, thus overcoming the attenuation of underwater acoustic signals. This active feature not only effectively improves the quality and SNR of the communication signal but also enables the protection of the SN’s location privacy by adding AN.

With the support of the existing UARIS architecture, the reflected and amplified signals of the $M$-element UARIS can be formulated as follows:
\begin{subequations}
	\begin{align}
		&\mathbf{y} = \mathbf{\Theta} \mathbf{x} + \mathbf{\Theta} \mathbf{v} + \mathbf{n_s},\ \mathbf{\Theta}=\text{Diag}(\boldsymbol{\theta^\top})\label{1a}\\
		&\boldsymbol{\theta} = \left[p_1 e^{j\theta_1},p_2 e^{j\theta_2},\cdots,p_M e^{j\theta_M}\right]^\top\in \mathbb{C}^{M\times 1}, \label{1b}
	\end{align}
\end{subequations}
where $\mathbf{\Theta}$ represents the reflection coefficient matrix of the UARIS, in which $p_m \in \mathbb{R}_{+}$ and $\theta_m$ denote the amplification factor and the phase shift factor of element $m$. Meanwhile, the active characteristics of UARIS introduce corresponding noise, which can be categorized into dynamic noise $\mathbf{\Theta} \mathbf{v}$ and static noise $\mathbf{n_s}$. Specifically, $\mathbf{v}$ is related to the input noise of UARIS and the inherent device noise, while $\mathbf{n_s}$ mainly represents the noise generated by the phase shift circuit. Since the energy of $\mathbf{n_s}$ is extremely small compared to $\mathbf{\Theta}\mathbf{v}$, $\mathbf{n_s}$ is neglected in the subsequent analysis. 

\begin{figure}[!t]
	\centering
	\includegraphics[width=3.2in]{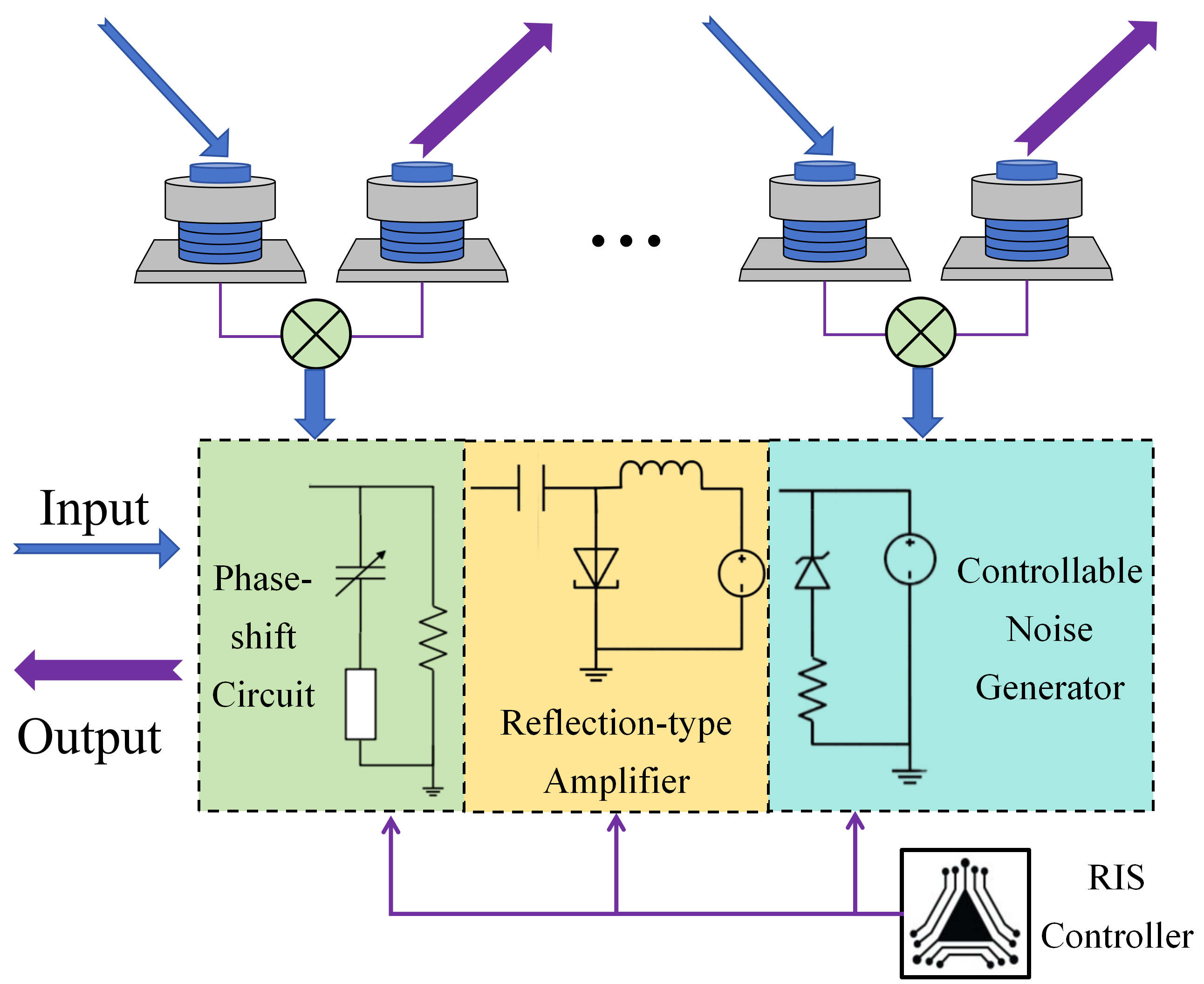}
	\caption{The hardware architectures of a UARIS with a controllable noise generator.}
	\label{UARIS}
\end{figure}

Additionally, the noise generated at the UARIS can be considered the AN to interfere with the ENs \cite{lyu2023robust}. To further protect the location privacy of the SN, we introduce a controllable noise generator to dynamically adjust the AN introduced at the UARIS, of which the architecture is shown in Fig. \ref{UARIS}. Specifically, based on the structure shown in Fig. \ref{UARIS}, we can model the AN $\mathbf{v}$ as $\mathbf{v}\sim\mathcal{CN}(\mathbf{0}_M, \eta \sigma_v^2\mathbf{I}_M)$, wherein $\eta$ denotes the noise factor of the AN generated at the UARIS.

%

\subsection{Underwater Acoustic Channel Model}
In this work, the primary research scenario is the acoustic propagation environment in shallow water areas. Different from wireless communications, the unique characteristics of underwater acoustic channels make the study of underwater signal transmission more complex and challenging. The attenuation pattern of underwater acoustic signals with respect to distance $d$ and frequency $f$ can be modeled as
\begin{equation}
	\Upsilon(d,f)=d^\epsilon\upsilon(f)^d,
\end{equation}
where $\epsilon$ denotes the propagation factor. According to \cite{jensen2011computational}, the absorption coefficient $\upsilon(f)$ can be obtained using Thorp's formula, which is formulated as
\begin{equation}
	10 \log_{10} \upsilon(f) = \frac{0.11 f^2}{1 + f^2} + \frac{44 f^2}{4100 + f^2} + \frac{2.75 f^2}{10^4} + 0.003.
\end{equation}

\subsection{Communication Model of the Receiving Node}
In this subsection, we will provide a detailed introduction to the application scenarios of UARIS-assisted UAC enhancement. Specifically, we consider a downlink UAC scenario and denote the symbol vector transmitted to $K$ RNs as $\mathbf{q}:=\left[q_1, q_2, \cdots, q_K\right]^\top \in \mathbb{C}^{K\times 1}$, which satisfies $\mathbb{E}\left\{\mathbf{q}\mathbf{q}^\dagger\right\}=\mathbf{I}_K$. As shown in Fig. \ref{FRAMEWORK}
, let $\mathbf{h}_k\in \mathbb{C}^{T\times 1}$, $\mathbf{H}\in \mathbb{C}^{M\times T}$ and $\mathbf{g}_k\in \mathbb{C}^{M\times 1}$ denote the channel matrix between the SN and RN $k$, that between the SN and the UARIS and that between the UARIS and RN $k$. Then, according to (\ref{1a}), signal $b_k\in\mathbb{C}$ received at RN $k$ can be modeled as
\begin{equation}\label{4}
	b_k = \left( \mathbf{h}_k^\dagger + \mathbf{g}_k^\dagger \mathbf{\Theta} \mathbf{H} \right) \sum_{i=1}^{K} \mathbf{w}_i q_i 
	+ \mathbf{g}_k^\dagger \mathbf{\Theta} \mathbf{v} + n_k
\end{equation}
where $\mathbf{w}_k\in\mathbb{C}^{T\times 1}$ denotes the beamforming vector designed by the SN for symbol $q_k$. Additionally, $\mathbf{g}_k^\dagger \mathbf{\Theta} \mathbf{v}$ denotes the AN introduced by the UARIS, while $n_k$ denote the background noise introduced at RN $k$, with $n_k\sim \mathcal{CN}(0, \sigma^2)$.
%

\subsection{Reception Signal Model of the Eavesdropping Node}
This subsection will introduce modeling the signals received by ENs in a shallow water environment. Specifically, assume that $J$ ENs possess spatial diversity and time synchronization characteristics, and each EN is equipped with an omni-directional hydrophone. Meanwhile, each EN records the received acoustic signals at predetermined observation intervals, then producing $N$ samples after sampling and baseband conversion. In addition, the scope of this work focuses on (approximately) constant-velocity environments and high-frequency signals, where the straight-line model is approximately valid.

Based on the three-ray model in \cite{weiss2022semi}, we added a transmission path through the UARIS, thus forming the four-ray model for shallow water transmission, illustrated in Fig. \ref{FRAMEWORK}. In this model, the four signal transmission paths are respectively:
\begin{enumerate}
	\item $L_{1j}$: The direct LOS path;
	\item $L_{2j}$: The sea surface reflected NLOS path;
	\item $L_{3j}$: The seabed reflected NLOS path;
	\item $L_{4j}$: The UARIS reflected NLOS path.
\end{enumerate}

According to Fig. \ref{FRAMEWORK}, the corresponding distances of these four paths can be further obtained, which are given by
\begin{subequations}
	\begin{align}
		&L_{1j}(\mathbf{p}_S) \triangleq \| \mathbf{p}_S - \mathbf{p}_j \|, \\
		&L_{2j}(\mathbf{p}_S) \triangleq \sqrt{l_{j}^2(\mathbf{p}_S) + (z + z_{j})^2}\\
		&L_{3j}(\mathbf{p}_S) \triangleq \sqrt{l_{j}^2(\mathbf{p}_S) + (2h - z - z_{j})^2}\\
		&L_{4j}(\mathbf{p}_S) \triangleq \| \mathbf{p}_S - \mathbf{p}_{U} \| + \| \mathbf{p}_{U} - \mathbf{p}_{j} \|
	\end{align}
\end{subequations}
where $h$ denotes the depth of the seabed, and $l_{j}(\mathbf{p}_S)$ represents the horizontal distance between the SN and EN $j$, which is defined as $l_{j}(\mathbf{p}_S)=\sqrt{(x_S-x_j)^2+(y_S-y_j)^2}$. We can further obtain the transmission delays for each path as follows:
\begin{equation}
	\tau_{ij}(\mathbf{p}_S)\triangleq \frac{L_{aj}(\mathbf{p}_S)}{c},\quad a = 1,2,3,4,\quad \forall j \in \mathcal{J},
\end{equation}
where $c$ represents the sound speed in shallow water.

Formally, the frequency domain representation $\mathbf{{u}}_j=\left[{u}_{j}[1],\cdots,{u}_{j}[N]\right]^\top\in\mathbb{C}^{N\times1}$ of the sampled baseband-converted signal obtained from EN $j$ can be modeled as
\begin{equation}\label{7}
    {u}_{j}[n] = \sum_{i=1}^{4} f_{ij} {s}[n] e^{-\jmath\omega_n \tau_{aj}(\mathbf{p}_S)} + {v}_{j}[n], \ n={1,\cdots N},
\end{equation}
where $\mathbf{{s}}=\left[{s}[1],\cdots,{s}[N]\right]^\top\in\mathbb{C}^{N\times1}$ denotes the frequency domain representation of the signal transmitted by the SN and $\omega_n=2\pi(n-1)/(NT_s)$ represents the angular frequency of the $k$-th frequency component. Additionally, let $\mathbf{h}_{E,j}\in \mathbb{C}^{T\times 1}$ and $\mathbf{g}_{E,j}\in \mathbb{C}^{M\times 1}$ denote the channel matrix between the SN and EN $j$, and that between the UARIS and EN $j$. Then, the noise ${v}_{j}[n]$ can be denoted as 
\begin{equation}\label{8}
    {v}_{j}[n] = \mathbf{g}_{E,j}^\dagger \mathbf{\Theta} \mathbf{v}[n] + {v}_{0}[n]
\end{equation}
where ${v}_{0}[n]$ is the background noise which satisfies ${v}_{0}[n]\sim \mathcal{CN}(0, \sigma^2)$.
Meanwhile, we can obtain the attenuation coefficients ${f_{ij}}$, which can be formulated as
\begin{subequations}
	\begin{align}
		f_{1j} &\triangleq \mathbf{h}_{E,j}^\dagger\sum_{i=1}^{K} \mathbf{w}_i, \quad \text{(LOS)} \\
		f_{2j} &\triangleq \Upsilon(L_{2j},f)\sum_{i=1}^{K} \mathbf{w}_i, \quad \text{(NLOS sea surface)} \\
		f_{3j} &\triangleq \kappa_b\Upsilon(L_{3j},f)\sum_{i=1}^{K} \mathbf{w}_i, \quad \text{(NLOS seabed)} \\
		f_{4j} &\triangleq \mathbf{g}_{E,j}^\dagger \mathbf{\Theta} \mathbf{H}\sum_{i=1}^{K} \mathbf{w}_i, \quad \text{(NLOS UARIS)}
	\end{align}
\end{subequations}
where $\kappa_b$ is the seabed reflection coefficient. 

For shorthand, we further define
\begin{subequations}
	\begin{align}
		&\mathbf{t}_{j,n}(\mathbf{p}_S) \triangleq \left[e^{-\jmath \omega_n \tau_{1j}(\mathbf{p}_S)},\cdots,e^{-\jmath \omega_k \tau_{4j}(\mathbf{p}_S)}\right]^\top\!\! 
		\in \mathbb{C}^{4 \times 1}, \\
		&\mathbf{T}_{j}(\mathbf{p}_S) \triangleq \left[\mathbf{t}_{j,1}(\mathbf{p}_S),\cdots,\mathbf{t}_{j,N}(\mathbf{p}_S)\right]^\top 
		\in \!\!\mathbb{C}^{N \times 4}, \\
		&\mathbf{f}_{j}\!\!  \triangleq \left[f_{1j},\cdots,f_{4j}\right]^\top\!\! 
		\in \mathbb{C}^{4 \times 1},\mathbf{G}_{j}=\text{Diag}(\mathbf{T}_{j}(\mathbf{p}_S)\mathbf{f}_{j}), \\
		&\mathbf{{v}}_j\triangleq\left[{v}_{j}[1],\cdots,{v}_{j}[N]\right]^\top\in\mathbb{C}^{N\times1}, \mathbf{U}_j \triangleq \text{Diag}(\mathbf{u}_j)
	\end{align}
\label{9}
\end{subequations}

According to \ref{9}, we can further simplify \ref{7} as
\begin{equation}\label{11}
	{\mathbf{u}}_{j} = \mathbf{G}_{j} {\mathbf{s}} + {\mathbf{v}}_{j} \in \mathbb{C}^{N \times 1}, \quad \forall j \in \mathcal{J}.
\end{equation}
\section{Semi-Blind Localization Method of the Eavesdropping Node}\label{SBL}
In this section, we introduce a localization method deployed at the ENs. Traditional underwater acoustic localization methods require a significant amount of prior knowledge, including channel information; however, such assumptions are difficult to establish in the eavesdropping localization model. Therefore, the ENs require a more robust localization method that do not rely on prior channel information to localize the SN, named Semi-Blind Localization (SBL) method \cite{weiss2022semi}. The following will provide a detailed analysis of this method and its Cram\'er-Rao Lower Bound (CRLB).
\subsection{Semi-Blind Localization Method}
To more accurately reflect the eavesdropping issue in a shallow open-sea environment, we assume the EN has no access to any information about the SN. Specifically, the EN has no prior knowledge of the channel matrix $\mathbf{F}\triangleq\left[\mathbf{f}_{1},\cdots,\mathbf{f}_{J}\right]\in\mathbb{C}^{4\times J}$ or the signal waveform transmitted by the SN. The EN only knows the seabed depth $h$ and the sound speed $c$. Without of generality, we can further assume that $\|{s}\|_2 = 1,$ viz., ${s} \in \mathcal{S}_N \triangleq \left\{ z \in \mathbb{C}^{N \times 1} : \|z\|_2 = 1 \right\}$, where $S_N$ denotes the $N$-dimensional unit sphere \cite{weiss2022semi}.

The solution of the SBL method can be regarded as the Maximum Likelihood Estimate (MLE) of $\mathbf{p}$, obtained through the joint estimation of all unknown deterministic model parameters. In this case, the maximum likelihood value of $\mathbf{p}$ is the solution to a nonlinear least squares problem:
\begin{equation}
	\hat{\mathbf{p}}_{e} \triangleq \mathop{\arg\min}\limits_{\mathbf{p} \in \mathbb{R}^{3 \times 1}} \min_{{\mathbf{s}} \in \mathcal{S}_N\atop\mathbf{F} \in \mathbb{C}^{4 \times J}} \sum_{j=1}^{J} \left\| {\mathbf{u}}_{j} - \operatorname{Diag}\left(\mathbf{T}_{j}(\mathbf{p})\mathbf{f}_{j}\right) {\mathbf{s}} \right\|^{2},
\end{equation}

Different from other traditional shallow sea localization methods, the SBL method we employ introduces some additional computational complexity but is more suited to practical eavesdropping localization scenarios and offers higher robustness. The method is summarized in Algorithm \ref{al2}, of which further analysis can be obtained in \cite{weiss2022semi}.
\begin{algorithm}[!t]
	\caption{Semi-Blind Localization Method.}
	\label{al2}
	\begin{algorithmic}[1]
		\STATE \textbf{Input:} Received signals $\left\{\mathbf{r}_j\right\}_{j=1}^J$, sound speed $c$, seabed depth $h$, grid resolution $\beta\in\mathbb{Z}_+$, the boundaries of the region of interest $\mathbf{x}_{range} = [x_{min}, x_{max}], \mathbf{y}_{range} = [y_{min}, y_{max}]$, and $\mathbf{z}_{range} = [z_{min}, z_{max}]$.
		\STATE \textbf{Output:} The estimated position of the SN $\hat{\mathbf{p}}_{e}$.
		\STATE Create a set of 3D grid points $\mathcal{X}\in\mathbb{R}^{\beta^3\times 3}$ based on the range of the 3D grid and the grid resolution,
		\FOR{$\mathbf{p}\in\mathcal{X}$ in parallel}
		\FOR{j = $1 \rightarrow J$}
		\STATE Compute the matrix $\mathbf{A}_j=\mathbf{T}_{j}^{\top} \mathbf{T}_{j}^{*} \in \mathbb{C}^{4 \times 4}$;
		\STATE Perform the Cholesky decomposition on matrix $\mathbf{A}$:
		\begin{equation}
			\mathbf{A}_{j} \triangleq \mathbf{\Omega}_{j}^{\mathrm{H}} \mathbf{\Omega}_{j} \in \mathbb{C}^{4 \times 4},
		\end{equation}
	    and obtain the matrix $\mathbf{\Omega}_{j}$;
	    \ENDFOR
	    \STATE Compute the matrix $\mathbf{Z}(\mathbf{p})$:
	    \begin{equation}
	    	\mathbf{Z}(\mathbf{p}) = \left[\mathbf{U}_1 \mathbf{T}_{1}^{*} \mathbf{\Omega}_{1}^{-1}, \cdots, \mathbf{U}_J \mathbf{T}_{J}^{*} \mathbf{\Omega}_{J}^{-1}\right]\in\mathbb{C}^{N\times4J}.
	    \end{equation}
        and construct the matrix $\mathbf{D}(\mathbf{p})$:
        \begin{equation}
        	\mathbf{D}(\mathbf{p}) \triangleq \mathbf{Z}(\mathbf{p})^\dagger \mathbf{Z}(\mathbf{p}) \in \mathbb{C}^{4J\times 4J}.
        \end{equation}
        \STATE Compute $\lambda_{\max}\left(\mathbf{D}(\mathbf{p})\right)$;
        \ENDFOR
        \STATE Find the optimal grid point $\mathbf{p}_o$:
        \begin{equation}
        	\hat{\mathbf{p}}_o =  \mathop{\arg\max}\limits_{\mathbf{p} \in \mathcal{P}} \lambda_{\max}\left(\mathbf{D}(\mathbf{p})\right).
        \end{equation}
        \STATE Starting with $\hat{\mathbf{p}}_o$ as the initial value, iteratively solve for $\hat{\mathbf{p}}_{e}$ using the trust-region method \cite{conn2000trust}.
	\end{algorithmic}
\end{algorithm}
\subsection{The Cram\'er-Rao Lower Bound Analysis for SBL}\label{CRLB}
In this subsection, we primarily analyze the Cramér-Rao Lower Bound (CRLB) for the SBL method. Specifically, without loss of generality, we focus on analyzing a spectrally flat waveform ${\mathbf{s}}$ \cite{weiss2022semi}. Therefore, $s[n]=\frac{1}{\sqrt{N}}e^{\jmath \varphi[n]}$ for all $n$, which means that the ENs can directly estimate the phase of the unknown source signal waveform. Meanwhile, since the EN also needs to estimate the unknown channel attenuation coefficient, the phase parameters to be estimated have only $N-1$ degrees of freedom, that is, the ENs only need to estimate $\boldsymbol{\varphi}\triangleq\left[\varphi[2],\cdots,\varphi[N]\right]^\top\in\mathbb{R}^{(N-1)\times1}$. 

To facilitate further derivation, we additionally define $\overline{\mathbf{u}}\triangleq[\mathbf{u}_1\cdots\mathbf{u}_J]^\top\in\mathbb{C}^{NJ\times 1}$, $\overline{\mathbf{G}}\triangleq[\mathbf{G}_1\cdots\mathbf{G}_J]^\top\in\mathbb{C}^{NJ\times N}$, and $\overline{\mathbf{v}}\triangleq[\mathbf{v}_1\cdots\mathbf{v}_J]^\top\in\mathbb{C}^{NJ\times 1}$. Thus, (\ref{11}) can be rewritten as
\begin{equation}
	\overline{\mathbf{u}} = \overline{\mathbf{G}}\mathbf{s}+\overline{\mathbf{v}} \in\mathbb{C}^{NJ\times 1}.
\end{equation}

Furthermore, according to (\ref{8}), we can obtain that $v_j[n]\sim\mathcal{CN}\left(0, \mathbf{g}_{E,j}^\dagger \mathbf{\Theta}\mathbf{\Theta}^\dagger\mathbf{g}_{E,j}\eta \sigma_v^2\right)$. Denoting $\boldsymbol{\sigma}^2_v\triangleq\eta \sigma_v^2\boldsymbol{\Gamma}\triangleq\eta \sigma_v^2\left[\Gamma_{v_1}\cdots\Gamma_{v_J}\right]^\top\in\mathbb{R}^{J\times1}_+$, wherein $\Gamma_{v_j}=\mathbf{g}_{E,j}^\dagger \mathbf{\Theta}\mathbf{\Theta}^\dagger\mathbf{g}_{E,j}+\sigma^2/(\eta \sigma_v^2)$, it follows that
\begin{equation}
	\overline{\mathbf{u}}\sim\mathcal{CN}(\overline{\mathbf{G}}\mathbf{s},\boldsymbol{\Sigma}),\ \boldsymbol{\Sigma}\triangleq\eta \sigma_v^2\text{Diag}(\boldsymbol{\Gamma})\otimes\mathbf{I}_N
\end{equation}

Thus, we can calculate the Fisher information matrix (FIM) elements for the Complex Normal (CN) signal model $\overline{\mathbf{u}}$, which can be formulated as \cite{collier2005fisher}
\begin{equation}
	\begin{aligned}
		&J[\vartheta_a, \vartheta_b] = \mathrm{Tr} \left( \boldsymbol{\Sigma}^{-1} \frac{\partial \boldsymbol{\Sigma}}{\partial \vartheta_a} \boldsymbol{\Sigma}^{-1} \frac{\partial \boldsymbol{\Sigma}}{\partial \vartheta_b} \right) \\
		&+ 2\mathcal{R} \left\{ \frac{\partial \left(\overline{\mathbf{G}}\mathbf{s}\right)^\dagger}{\partial \vartheta_a} \boldsymbol{\Sigma}^{-1} \frac{\partial \left(\overline{\mathbf{G}}\mathbf{s}\right)}{\partial \vartheta_b} \right\},
		\ \forall a, b \in \{1, \ldots, N_\vartheta\},
	\end{aligned}
\end{equation}
where let $\boldsymbol{\vartheta}$ denote the vector of all unknown deterministic parameters that the ENs need to estimate, which is defined as
\begin{equation}
	\boldsymbol{\vartheta} = \left[\mathbf{p}, \boldsymbol{\varphi}^\top, \mathcal{R}\left\{\text{vec}(\mathbf{F})\right\},\mathcal{I}\left\{\text{vec}(\mathbf{F})\right\},(\boldsymbol{\sigma}^2_v)^\top\right]\in\mathbb{R}^{1\times N_\vartheta}
\end{equation}
where $N_\vartheta=3+(N-1)+2\times4J+J$ represents the number of parameters to be estimated, and we define $\mathbf{J}(\boldsymbol{\vartheta})$ as the FIM.

In the received signal model of the ENs, the parameters to be estimated differ between the mean vector and the covariance matrix. Therefore, we can further deduce that
\begin{subequations}
	\begin{align}
		&J\left[\sigma_{v_a}^2,\sigma_{v_b}^2\right]=N\delta_{ab},\ \forall a,b\in\mathcal{J},\\
		&J\left[\sigma_{v_a}^2,\vartheta_a\right]=0, \ a=1,\cdots,N_\vartheta-J.
	\end{align}
\end{subequations}
namely the FIM has a block diagonal structure. Consequently, the elements in $\mathbf{J}(\boldsymbol{\vartheta})$ corresponding to the estimated parameters related to the signal model can be calculated as
\begin{equation}\label{22}
	J[\vartheta_a, \vartheta_b]=2\mathcal{R} \left\{ \frac{\partial \left(\overline{\mathbf{G}}\mathbf{s}\right)^\dagger}{\partial \vartheta_a} \boldsymbol{\Sigma}^{-1} \frac{\partial \left(\overline{\mathbf{G}}\mathbf{s}\right)}{\partial \vartheta_b} \right\}
\end{equation}

To obtain the complete FIM, we also need to compute the derivative of $\overline{\mathbf{G}}\mathbf{s}$ with respect to the parameters $\boldsymbol{\vartheta}$, which can be formulated as for all $i\in\{1,\cdots,4\}$, $j\in\mathcal{J}$, and $\vartheta_p\in\{x_p,y_p,z_p\}$
\begin{equation}\label{23}
	\frac{\partial \overline{\mathbf{G}} \mathbf{s}}{\partial \vartheta_p} = \frac{\partial}{\partial \vartheta_p} 
	\begin{bmatrix} 
		\mathrm{Diag}(\mathbf{T}_{1}\mathbf{f}_{1}) \\ 
		\vdots \\ 
		\mathrm{Diag}(\mathbf{T}_{J}\mathbf{f}_{J}) 
	\end{bmatrix} 
	\mathbf{s} 
	= \begin{bmatrix} 
		\mathrm{Diag}\left(\frac{\partial \mathbf{T}_{1}}{\partial \vartheta_p} \mathbf{f}_{1} \right) \\ 
		\vdots \\ 
		\mathrm{Diag}\left(\frac{\partial \mathbf{T}_{J}}{\partial \vartheta_p} \mathbf{f}_{J} \right) 
	\end{bmatrix} 
	\mathbf{s}.
\end{equation}
\begin{equation}
	\frac{\partial \overline{\mathbf{G}} \mathbf{s}}{\partial \varphi[n]}= \overline{\mathbf{G}} \left( \sqrt{\frac{1}{N}} \cdot \frac{\partial e^{\jmath\varphi[n]}}{\partial \varphi[n]} \right) =(\jmath \sqrt{\frac{1}{N}} e^{\jmath\varphi[n]}) \overline{\mathbf{G}} \mathbf{e}_n.
\end{equation}
\begin{equation}
	\frac{\partial \overline{\mathbf{G}} \mathbf{s}}{\partial \mathcal{R}\{f_{ij}\}} 
	= \left( \mathbf{e}_j \otimes \frac{\partial \mathrm{Diag}(\mathbf{T}_{j}\mathbf{f}_{j})}{\partial \mathcal{R}\{f_{ij}\}} \right) \mathbf{s}  
	= \mathbf{e}_j \otimes (\mathrm{Diag}(\mathbf{T}_j \mathbf{e}_{i}) \mathbf{s}).
\end{equation}
\begin{equation}\label{26}
	\frac{\partial \overline{\mathbf{G}} \mathbf{s}}{\partial \mathcal{I}\{f_{ij}\}} 
	\!\!=\!\! \jmath\mathbf{e}_j \otimes (\mathrm{Diag}(\mathbf{T}_j \mathbf{e}_{i})\mathbf{s}).
\end{equation}

By substituting (\ref{23})-(\ref{26}) into (\ref{22}), the complete FIM $\mathbf{J}(\boldsymbol{\vartheta})$ can be computed. Furthermore, according to (\ref{22}), it can be observed that the FIM block related to the signal components is inversely proportional to the noise $\eta \sigma_v^2$ introduced by the UARIS. Therefore, the CRLB block associated with the signal components is directly proportional to $\eta \sigma_v^2$. Finally, the CRLB for this signal estimation model can be given by
\begin{subequations}\label{27}
	\begin{align}
		&\mathbb{E} \left[ \left(\hat{\boldsymbol{\vartheta}} - \boldsymbol{\vartheta}\right)\left(\hat{\boldsymbol{\vartheta}} - \boldsymbol{\vartheta}\right)^\top \right] \succeq \mathbf{J}^{-1}(\boldsymbol{\vartheta}) \triangleq \mathrm{CRLB}(\boldsymbol{\vartheta}),\\
		&\implies \mathrm{MSE}(\hat{\mathbf{p}}, \mathbf{p}) \geq \sum_{i=1}^{3} \left[\mathrm{CRLB}(\boldsymbol{\vartheta})_{i,i}\right],
	\end{align}
\end{subequations}

\section{Joint Transmit Bemaforming and Reflect Precoding Design with Artificial Noise}\label{opt}
To investigate the communication enhancement and location privacy protection supported by the underwater acoustic RIS with AN sources in underwater communication scenarios, we will model a multi-objective optimization problem for this scenario in this section. Furthermore, a joint optimization scheme for transmit beamforming, reflective precoding, and noise factor is proposed.
\subsection{Problem Formulation}
According to the underwater communication model in (\ref{4}), the signal-to-interference-plus-noise ratio (SINR) at RN $k$ can be calculated as
\begin{subequations}
	\begin{align}
		&\gamma_k = \frac{|\overline{\mathbf{h}}_k^\dagger \mathbf{w}_k|^2}{\sum_{a=1, a \neq k}^{K} |\overline{\mathbf{h}}_k^\dagger \mathbf{w}_a|^2 + \left\| \mathbf{g}_k^\dagger \mathbf{\Theta} \right\|^2 \eta\sigma_v^2 + \sigma^2},\\
		&\overline{\mathbf{h}}_k^\dagger=\mathbf{h}_k^\dagger + \mathbf{g}_k^\dagger \mathbf{\Theta} \mathbf{H}=\mathbf{h}_k^\dagger +\boldsymbol{\theta}^\top\text{Diag}(\mathbf{g}_k^\dagger)\mathbf{H}\in\mathbb{C}^{1\times T}
	\end{align}
\end{subequations}
where $\overline{\mathbf{h}}_k^\dagger=\mathbf{h}_k^\dagger + \mathbf{g}_k^\dagger \mathbf{\Theta} \mathbf{H}\in\mathbb{C}^{1\times T}$ represents the equivalent channel vector between the SN and RN $k$. 

Furthermore, to jointly optimize the UARIS-assisted communication enhancement and location privacy protection, we formulate the following optimization problem:
\begin{subequations}\label{p1}
	\begin{align}
		\mathcal{P}_1 : \max_{\mathbf{w}, \mathbf{\Theta}, \eta} &R_1 = \sum_{k=1}^{K} \log_2 (1 + \gamma_k) + \sum_{i=1}^{3} |\mathrm{CRLB}(\boldsymbol{\vartheta})_{i,i}|, \\
		\text{s.t.} \ &\mathcal{C}_1 : \sum_{k=1}^{K} \|\mathbf{w}_k\|^2 \leq P_{S}^{\max},\\
		&\mathcal{C}_2 : \sum_{k=1}^{K} \|\mathbf{\Theta} \mathbf{G} \mathbf{w}_k\|^2 + \|\mathbf{\Theta}\|_F^2 \eta\sigma_v^2 \leq P_U^{\max},\\
		&\mathcal{C}_3 : \eta\geq1.
	\end{align}
\end{subequations}
where $\mathbf{w}\triangleq\left[\mathbf{w}_1^\top,\cdots,\mathbf{w}_K^\top\right]^\top\in\mathbb{C}^{TK\times1}$.

According to the CRLB analysis in Subsection \ref{CRLB}, since the $\sum_{i=1}^{3} |\mathrm{CRLB}(\boldsymbol{\vartheta})_{i,i}|$ is directly proportional to the controllable noise $\eta\sigma_v^2$ introduced by the UARIS, the optimization problem (\ref{p1}) can be reformulated as
\begin{subequations}\label{p2}
	\begin{align}
		\mathcal{P}_2 : \max_{\mathbf{w}, \mathbf{\Theta}, \eta} &R_2 = \sum_{k=1}^{K} \left[\log_2 (1 + \gamma_k) - \frac{\xi_0}{K}\log_2(1+\frac{1}{\eta\sigma_v^2})\right] , \\
		\text{s.t.} \ &\mathcal{C}_1 : \sum_{k=1}^{K} \|\mathbf{w}_k\|^2 \leq P_{S}^{\max},\\
		&\mathcal{C}_2 : \sum_{k=1}^{K} \|\mathbf{\Theta} \mathbf{G} \mathbf{w}_k\|^2 + \|\mathbf{\Theta}\|_F^2 \eta\sigma_v^2 \leq P_U^{\max},\\
		&\mathcal{C}_3 : \eta\geq1,
	\end{align}
\end{subequations}
where $\xi$ represents the weight parameter that balances communication enhancement and location privacy protection, and we further define $\xi=\xi_0/K\ll1$. Moreover, for sufficiently large $\gamma_k$ and $1/\eta\sigma_v^2$, $\log_2 \left[\frac{1 + \gamma_k}{(1+1/\eta\sigma_v^2)^{\xi}}\right]$ can be approximated by $\log_2 \left[1 + \gamma_k(\eta\sigma_v^2)^{\xi}\right]$. Thus, the optimization problem (\ref{p2}) can be reformulated as
\begin{subequations}\label{p3}
	\begin{align}
		\mathcal{P}_3 : \max_{\mathbf{w}, \mathbf{\Theta}, \eta} &R_3 = \sum_{k=1}^{K} \log_2 \left[1 + \gamma_k(\eta\sigma_v^2)^{\xi}\right] , \\
		\text{s.t.} \ &\mathcal{C}_1 : \sum_{k=1}^{K} \|\mathbf{w}_k\|^2 \leq P_{S}^{\max},\\
		&\mathcal{C}_2 : \sum_{k=1}^{K} \|\mathbf{\Theta} \mathbf{G} \mathbf{w}_k\|^2 + \|\mathbf{\Theta}\|_F^2 \eta\sigma_v^2 \leq P_U^{\max},\\
		&\mathcal{C}_3 : \eta\geq1.
	\end{align}
\end{subequations}

The non-convexity of (\ref{p3}) makes it challenging to solve directly. To handle this non-convex logarithmic and fractional problem, we employ the Fractional Programming (FP) method to decouple (\ref{p3}), enabling the optimization of multiple variables separately. Therefore, the following lemma needs to be introduced \cite{shen2018fractional}:

\begin{lemma}\label{l1}
	By introducing auxiliary variables $\boldsymbol{\zeta}\triangleq \left[\zeta_1,\cdots,\zeta_K\right]$ and $\boldsymbol{\chi}\triangleq \left[\chi_1,\cdots,\chi_K\right]$ based on (\ref{p3}), it can be equivalently transformed into
	\begin{subequations}\label{p4}
		\begin{align}
			\mathcal{P}_4 : &\max_{\mathbf{w}, \mathbf{\Theta}, \eta, \boldsymbol{\zeta}, \boldsymbol{\chi}} R_4 = \sum_{k=1}^{K} \left[\log_2 \left(1 + \zeta_k\right)-\zeta_k\right]\notag\\	
			+\sum_{k=1}^{K}&\left[2\sqrt{(\eta\sigma_v^2)^{\xi}(1+\zeta_k)} \mathcal{R}\left\{\chi^*\overline{\mathbf{h}}_k^\dagger \mathbf{w}_k\right\} \right]\notag\\
			-\sum_{k=1}^{K}& |\chi_k|^2\left(\sum_{a=1}^{K} |\overline{\mathbf{h}}_k^\dagger \mathbf{w}_a|^2 + \left\| \mathbf{g}_k^\dagger \mathbf{\Theta} \right\|^2 \eta\sigma_v^2 + \sigma^2\right)\\
			\text{s.t.} \ &\mathcal{C}_1 : \sum_{k=1}^{K} \|\mathbf{w}_k\|^2 \leq P_{S}^{\max},\label{c1}\\
			&\mathcal{C}_2 : \sum_{k=1}^{K} \|\mathbf{\Theta} \mathbf{G} \mathbf{w}_k\|^2 + \|\mathbf{\Theta}\|_F^2 \eta\sigma_v^2 \leq P_U^{\max},\label{c2}\\
			&\mathcal{C}_3 : \eta\geq1. \label{c3}
		\end{align}
	\end{subequations}
\end{lemma}

$\mathit{Proof.}$ Detailed proof can be found in \cite{shen2018fractional}.
\subsection{Joint Optimization Scheme for Beamforming, Reflection Matrix, and Noise Factor}
Based on Lemma \ref{l1}, the original joint optimization problem can be decoupled into an alternate optimization of the SN beamforming vector $\mathbf{w}$, UARIS reflection matrix $\boldsymbol{\Theta}$, noise factor $\eta$, auxiliary variables $\boldsymbol{\zeta}$ and $\boldsymbol{\chi}$. According to \cite{shen2018fractional}, if all variables in Problem 1 are optimal in each iteration update, a locally optimal solution to (\ref{p4}) can be obtained after convergence. Therefore, we will provide a detailed introduction to the optimization steps for each variable, and summarize the proposed joint optimization scheme in Algorithm \ref{al3}.
\begin{algorithm}[!t]
	\caption{Joint Optimization Scheme for Beamforming, Reflection Matrix, and Noise Factor.}
	\label{al3}
	\begin{algorithmic}[1]
		\STATE \textbf{Input:} Channel matrices $\mathbf{h}_k\in \mathbb{C}^{T\times 1}$, $\mathbf{H}\in \mathbb{C}^{M\times T}$ and $\mathbf{g}_k\in \mathbb{C}^{M\times 1}$ for all $k \in \mathcal{K}$.
		\STATE \textbf{Output:} The SN beamforming vector $\mathbf{w}$, UARIS reflection matrix $\boldsymbol{\Theta}$, and noise factor $\eta$,
		\STATE Initialize $\mathbf{w}$, $\boldsymbol{\Theta}$ and $\eta=1$.
		\WHILE{no convergence of $R_3$}
		\STATE Fix variables $\left(\mathbf{w},\boldsymbol{\Theta},\eta,\boldsymbol{\chi}\right)$, and update auxiliary variables $\boldsymbol{\zeta}$ by (\ref{zeta});
		\STATE Fix variables $\left(\mathbf{w},\boldsymbol{\Theta},\eta,\boldsymbol{\zeta}\right)$, and update auxiliary variables $\boldsymbol{\chi}$ by (\ref{chi});
		\STATE Fix variables $\left(\mathbf{w},\boldsymbol{\Theta},\boldsymbol{\zeta},\boldsymbol{\chi}\right)$, and update the noise factor $\eta$ by (\ref{eta});
		\STATE Fix variables $\left(\boldsymbol{\Theta},\eta,\boldsymbol{\zeta},\boldsymbol{\chi}\right)$, and use CVX tool to update the SN beamforming vector $\mathbf{w}$ by (\ref{w});
		\STATE Fix variables $\left(\mathbf{w},\eta,\boldsymbol{\zeta},\boldsymbol{\chi}\right)$, and use a binary search to obtain the optimal Lagrange multiplier $\mu$ to update the UARIS reflection matrix $\boldsymbol{\Theta}$ by (\ref{theta})
		\ENDWHILE
	\end{algorithmic}
\end{algorithm}
\begin{enumerate}
	\item Optimize $\boldsymbol{\zeta}$: Fix variables $\left(\mathbf{w},\boldsymbol{\Theta},\eta,\boldsymbol{\chi}\right)$, and optimize auxiliary variables $\boldsymbol{\zeta}$ as
	\begin{equation}\label{zeta}
		\zeta_k^{op} = \frac{|\overline{\mathbf{h}}_k^\dagger \mathbf{w}_k|^2(\eta\sigma_v^2)^{\xi}}{\sum_{a=1}^{K} |\overline{\mathbf{h}}_k^\dagger \mathbf{w}_a|^2 + \left\| \mathbf{g}_k^\dagger \mathbf{\Theta} \right\|^2 \eta\sigma_v^2 + \sigma^2},
	\end{equation} 
    \item Optimize $\boldsymbol{\chi}$: Fix variables $\left(\mathbf{w},\boldsymbol{\Theta},\eta,\boldsymbol{\zeta}\right)$, and optimize auxiliary variables $\boldsymbol{\chi}$ by solving $\partial R_4 / \partial \chi_k=0$ as
    \begin{equation}\label{chi}
    	\chi_k^{op} = \frac{\sqrt{(\eta\sigma_v^2)^{\xi}(1+\zeta_k)}\overline{\mathbf{h}}_k^\dagger \mathbf{w}_k}{\sum_{a=1}^{K} |\overline{\mathbf{h}}_k^\dagger \mathbf{w}_a|^2 + \left\| \mathbf{g}_k^\dagger \mathbf{\Theta} \right\|^2 \eta\sigma_v^2 + \sigma^2},
    \end{equation}
    \item Optimize $\eta$: Fix variables $\left(\mathbf{w},\boldsymbol{\Theta},\boldsymbol{\zeta},\boldsymbol{\chi}\right)$, and optimize the noise factor $\eta$ by solving $\partial R_4 / \partial \eta=0$ as
    \begin{subequations}\label{eta}
    	\begin{align}
    		&\eta^* = \left(\frac{2}{\xi}\frac{\sum_{k=1}^{K}|\chi_k|^2\left\| \mathbf{g}_k^\dagger \mathbf{\Theta} \right\|^2 \sigma_v^2}{\sum_{k=1}^{K}2\sqrt{\sigma_v^{2\xi}(1+\zeta_k)}\mathcal{R}\left\{\chi^*\overline{\mathbf{h}}_k^\dagger \mathbf{w}_k\right\}}\right)^{\frac{2}{\xi-2}},\\
    		&\eta^{op}=\max\left(1,\min\left(\eta^*,\frac{P_U^{\max}-\sum_{k=1}^{K} \|\mathbf{\Theta} \mathbf{G} \mathbf{w}_k\|^2}{\|\mathbf{\Theta}\|_F^2\sigma_v^2}\right)\right)
    	\end{align}
    \end{subequations}
    \item Optimize $\mathbf{w}$: For shorthand, we further define
    \begin{subequations}\label{w}
    	\begin{align}
    		&\boldsymbol{\alpha}_k \!\!=\!\! \sqrt{(\eta\sigma_v^2)^{\xi}(1\!+\!\zeta_k)}\chi_k\overline{\mathbf{h}}_k^\dagger, \boldsymbol{\alpha}\!\!\triangleq\!\!\left[\boldsymbol{\alpha}_1^\top\!,\!\cdots\!,\!\boldsymbol{\alpha}_1^\top\right]^\top\!\!\!,\\
    		&\mathbf{B}\!\!=\!\!\mathbf{I}_K\otimes\sum_{k=1}^K |\chi_k|^2 \overline{\mathbf{h}}_k\overline{\mathbf{h}}_k^\dagger, \mathbf{C}\!\!=\!\!\mathbf{I}_K\otimes(\mathbf{H}^\dagger\boldsymbol{\Theta}^\dagger\boldsymbol{\Theta}\mathbf{H})
    	\end{align}
    \end{subequations}
    
    Then, fix variables $\left(\boldsymbol{\Theta},\eta,\boldsymbol{\zeta},\boldsymbol{\chi}\right)$, and obtain a new optimization problem for the SN beamforming vector $\mathbf{w}$ based on (\ref{p4}), which can be formulated as
    \begin{subequations}\label{p5}
    	\begin{align}
    		\mathcal{P}_5 : \max_{\mathbf{w}}\  &\mathcal{R}\left\{2\boldsymbol{\alpha}^\dagger\mathbf{w}\right\}-\mathbf{w}^\dagger\mathbf{B}\mathbf{w}, \\
    		\text{s.t.} \ &\mathcal{C}_1 : \|\mathbf{w}\|^2 \leq P_{S}^{\max},\\
    		&\mathcal{C}_2 : \mathbf{w}^\dagger\mathbf{C}\mathbf{w} \leq P_U^{\max}-\|\mathbf{\Theta}\|_F^2 \eta\sigma_v^2,
    	\end{align}
    \end{subequations} 

    As a standard quadratic constraint quadratic programming, optimization problem (\ref{p5}) can be solved directly using CVX tool to obtain the optimal beamforming vector $\mathbf{w}^{op}$.
    \item Optimize $\boldsymbol{\Theta}$: For shorthand, we further define
    \begin{subequations}
    	\begin{align}
    		\mathbf{v} = &\sum_{k=1}^{K} \sqrt{(\eta\sigma_v^2)^{\xi}(1\!+\!\zeta_k)} \mathrm{Diag} \left( \boldsymbol{\chi}_k^* \mathbf{g}_k^\dagger \right) \mathbf{H} \mathbf{w}_k  \notag\\
    		&-\sum_{k=1}^{K} |\boldsymbol{\chi}_k|^2 \mathrm{Diag} (\mathbf{g}_k^\dagger) \mathbf{H} \sum_{a=1}^{K} \mathbf{w}_a \mathbf{w}_a^\dagger \mathbf{h}_k,\\
    		\boldsymbol{\Lambda} = &\sum_{k=1}^{K} |\boldsymbol{\chi}_k|^2 \sum_{j=1}^{K} \mathrm{Diag} \left( \mathbf{g}_k^\dagger \right) \mathbf{H} \mathbf{w}_j \mathbf{w}_j^\dagger \mathbf{H}^\dagger \mathrm{Diag} \left( \mathbf{g}_k \right) \notag\\
    		&+\sum_{k=1}^{K} |\boldsymbol{\chi}_k|^2 \mathrm{Diag} \left( \mathbf{g}_k^\dagger \right) \mathrm{Diag} \left( \mathbf{g}_k \right) \sigma_v^2,\\
    		\boldsymbol{\Psi} = &\sum_{k=1}^{K} \mathrm{Diag}(\mathbf{H} \mathbf{w}_k) \left( \mathrm{Diag}(\mathbf{H} \mathbf{w}_k) \right)^\dagger + \eta\sigma_v^2 \mathbf{I}_M.
    	\end{align}
    \end{subequations}

    Then, fix variables $\left(\mathbf{w},\eta,\boldsymbol{\zeta},\boldsymbol{\chi}\right)$, and obtain a new optimization problem for the UARIS reflection matrix $\boldsymbol{\Theta}$ based on (\ref{p4}), which can be formulated as
    \begin{subequations}\label{p6}
    	\begin{align}
    		\mathcal{P}_6 : \max_{\boldsymbol{\theta}}\  &\mathcal{R}\left\{2\boldsymbol{\theta}^\dagger\mathbf{v}\right\}-\boldsymbol{\theta}^\dagger\boldsymbol{\Lambda}\boldsymbol{\theta}, \\
    		\text{s.t.} \ &\mathcal{C}_1 : \boldsymbol{\theta}^\dagger\boldsymbol{\Psi}\boldsymbol{\theta} \leq P_U^{\max}.\label{con}
    	\end{align}
    \end{subequations}
 
     Furthermore, we use the Lagrange multiplier method to solve optimization problem (\ref{p6}). Specifically, we introduce the Lagrange multiplier $\mu$ and construct the Lagrange function as 
     \begin{equation}
     	\mathcal{L}(\boldsymbol{\theta},\mu)=\mathcal{R}\left\{2\boldsymbol{\theta}^\dagger\mathbf{v}\right\}-\boldsymbol{\theta}^\dagger\boldsymbol{\Lambda}\boldsymbol{\theta}+\mu(P_U^{\max}-\boldsymbol{\theta}^\dagger\boldsymbol{\Psi}\boldsymbol{\theta}).
     \end{equation}
 
     Next, take the derivative of the Lagrange function $\mathcal{L}(\boldsymbol{\theta},\mu)$ with respect to $\boldsymbol{\theta}$ and set the derivative to zero:
     \begin{equation}\label{l}
     	\frac{\partial \mathcal{L}}{\partial \boldsymbol{\theta}} = 2 \mathbf{v} - 2 \boldsymbol{\Lambda} \boldsymbol{\theta} - 2 \mu \boldsymbol{\Psi} \boldsymbol{\theta} = 0.
     \end{equation}
     
     By solving the linear equation system \ref{l}, the solution for $\boldsymbol{\theta}$ can be obtained as
     \begin{equation}\label{theta}
     	\boldsymbol{\theta}=\left(\boldsymbol{\Lambda}+\mu \boldsymbol{\Psi}\right)^{-1}\mathbf{v},
     \end{equation}
     wherein the optimal Lagrange multiplier $\mu$ that satisfies power constraint (\ref{con}) can be obtained through a binary search \cite{boyd2011distributed}.
\end{enumerate}
\subsection{Convergence Analysis}
In this subsection, we will conduct a convergence analysis of the proposed joint optimization algorithm. Specifically, Algorithm $\ref{al3}$ decouples the joint optimization problem of beamforming, reflection matrix, and noise factors into several subproblems. Since the updates in each iteration of Algorithm $\ref{al3}$ are optimal solutions to these subproblems, the algorithm can converge to a local optimum after multiple iterations. A detailed proof will be provided next.

To facilitate the proof, we introduce the superscript $t$ for each variable to denote its value in the $t$-th iteration. Then, the convergence of Algorithm $\ref{al3}$ can be formulated as
\begin{equation}
	\begin{aligned}
		&R_4\left(\mathbf{w}^{t+1}, \mathbf{\Theta}^{t+1}, \eta^{t+1}, \boldsymbol{\zeta}^{t+1}, \boldsymbol{\chi}^{t+1}\right)\overset{(\text{i})}{\geq}\\
		&R_4\left(\mathbf{w}^{t+1}, \mathbf{\Theta}^{t}, \eta^{t+1}, \boldsymbol{\zeta}^{t+1}, \boldsymbol{\chi}^{t+1}\right)\overset{(\text{ii})}{\geq}\\
		&R_4\left(\mathbf{w}^{t}, \mathbf{\Theta}^{t}, \eta^{t+1}, \boldsymbol{\zeta}^{t+1}, \boldsymbol{\chi}^{t+1}\right)\overset{(\text{iii})}{\geq}\\
		&R_4\left(\mathbf{w}^{t}, \mathbf{\Theta}^{t}, \eta^{t}, \boldsymbol{\zeta}^{t+1}, \boldsymbol{\chi}^{t+1}\right)\overset{(\text{iv})}{\geq}\\
		&R_4\left(\mathbf{w}^{t}, \mathbf{\Theta}^{t}, \eta^{t}, \boldsymbol{\zeta}^{t+1}, \boldsymbol{\chi}^{t}\right)\overset{(\text{v})}{\geq}R_4\left(\mathbf{w}^{t}, \mathbf{\Theta}^{t}, \eta^{t}, \boldsymbol{\zeta}^{t}, \boldsymbol{\chi}^{t}\right)
	\end{aligned}
\end{equation}
wherein the validity of (i) and (ii) holds because, in the $t$-th iteration, $\mathbf{w}^{t+1}$ and $\mathbf{\Theta}^{t+1}$ are the optimal solutions to subproblems (\ref{p5}) and (\ref{p6}), respectively, in that iteration. Meanwhile, (iii), (iv) and (v) follow since in the $t$-th iteration, $\eta^{t+1}$, $\boldsymbol{\zeta}^{t+1}$, and $\boldsymbol{\chi}^{t+1}$ are all updates that maximize $R_4$. Therefore, the optimization objective $R_4$ is monotonically non-decreasing in each iteration. Additionally, due to the constraints in (\ref{c1}), (\ref{c2}) and (\ref{c3}), the value of $R_4$ is upper-bounded, which ensures that Algorithm \ref{al3} will converge to a local optimum.
\section{Simulation Results and Analysis}\label{sim}
In this subsection, we will conduct a simulation to validate the proposed joint optimization scheme. Specifically, in Subsection \ref{vacrlb}, we validate the CRLB analysis provided in (\ref{27}). In Subsection \ref{cov}, simulation results are presented to evaluate the coverage performance of the UARIS-assisted underwater communication system with AN. In Subsection \ref{eff} and \ref{effe}, we discuss the impact of optimized weight $\xi$ on the performance optimization of the proposed scheme. Finally, in Subsection \ref{effe2}, the impact of the number of UARIS elements on optimization performance is explored from the hardware perspective.
\subsection{Simulation Setting}
 In the simulation experiments, we assume the presence of four legitimate receiving nodes, all evenly distributed on a spherical surface with point $\mathbf{p}_{R}^0 \triangleq\left[x_{R}^0\ y_{R}^0 \ z_{R}^0\right]^\top$ as the center of the sphere. Meanwhile, let $d_{SR} \triangleq \| \mathbf{p}_S - \mathbf{p}_{R}^0 \|$ denote the average distance between the SN and the RNs, and $d_{ER}$ represents the average distance between the ENs and the RNs. The other simulation parameters are summarized in Table \ref{t1}. Additionally, to ensure the fairness of the comparative experiments, we define a total power limit $P_t^{max}$. Specifically, for the UARIS-assisted UAC system, the total power is given by $P_t^{max} = P_S^{max} + P_U^{max}$, wherein $P_S^{max} = 0.9P_t^{max}$ and $P_U^{max} = 0.1P_t^{max}$. For the UAC system without UARIS, the total power is given by $P_t^{max} = P_S^{max}$. To demonstrate the effectiveness of the proposed joint optimization scheme, we conduct simulations for the following three schemes:
\begin{enumerate}
	\item \textbf{Without UARIS (M1):} In an underwater communication system without UARIS, the WMMSE algorithm proposed in \cite{shi2011iteratively} is used to optimize the beamforming of the SN.
	\item \textbf{UARIS with fixed noise (M2):} In a UARIS-assisted underwater communication system with fixed noise, the algorithm proposed in \cite{zhang2022active} is used to jointly optimize the beamforming of the SN and the reflection matrix of the UARIS.
	\item \textbf{UARIS with AN  (M3):} In a UARIS-assisted underwater communication system with AN, the proposed Algorithm \ref{al3} is used to jointly optimize the beamforming of the SN, the reflection matrix of the UARIS and the noise factor.
\end{enumerate}
\begin{figure*}[t]
	\centering
	\subfigure[\label{fig4a}The 95\%-confidence ellipsoid without UARIS]{
		\begin{minipage}[t]{0.28\textwidth}
			\centering
			\includegraphics[width=2.3in,center]{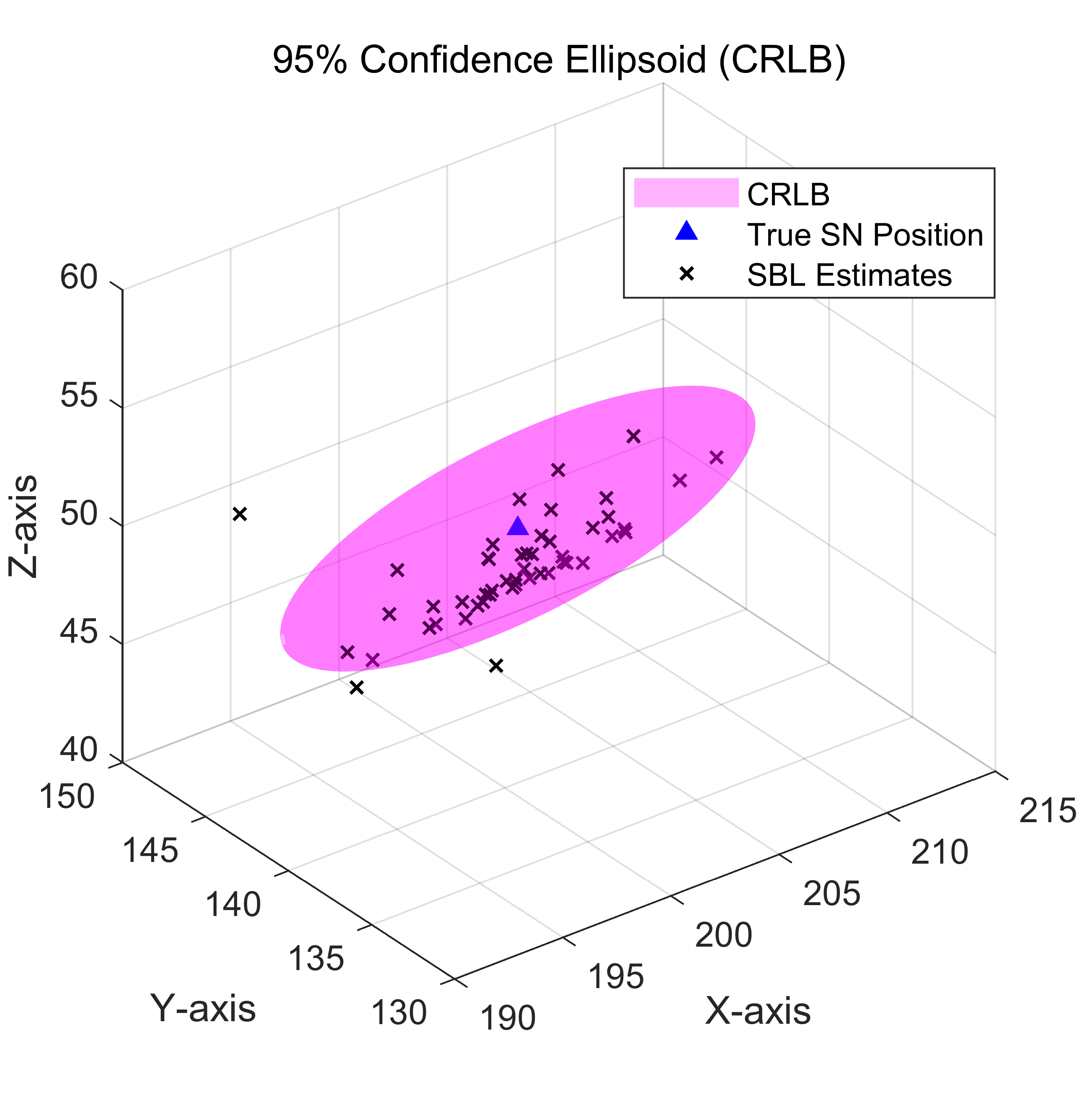}\\
		\end{minipage}%
	}\hspace{5.8mm}
	\centering
	\subfigure[\label{fig4b}The 95\%-confidence ellipsoid of UARIS with fixed noise]{
		\begin{minipage}[t]{0.28\textwidth}
			\centering
			\includegraphics[width=2.3in,center]{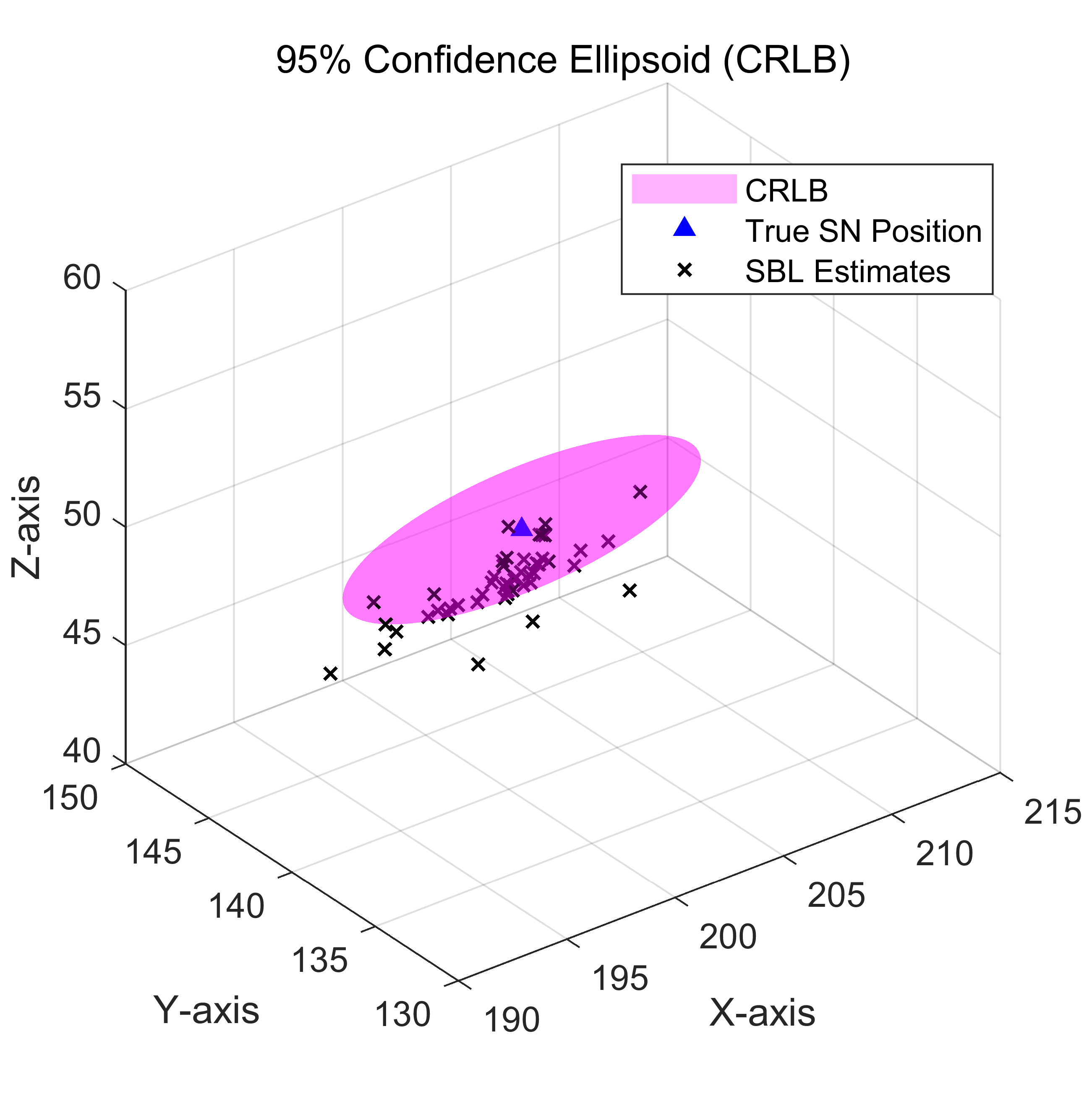}\\
		\end{minipage}%
	}\hspace{5.8mm}
	\centering
	\subfigure[\label{fig4c}The 95\%-confidence ellipsoid of UARIS with AN]{
		\begin{minipage}[t]{0.28\textwidth}
			\centering
			\includegraphics[width=2.3in,center]{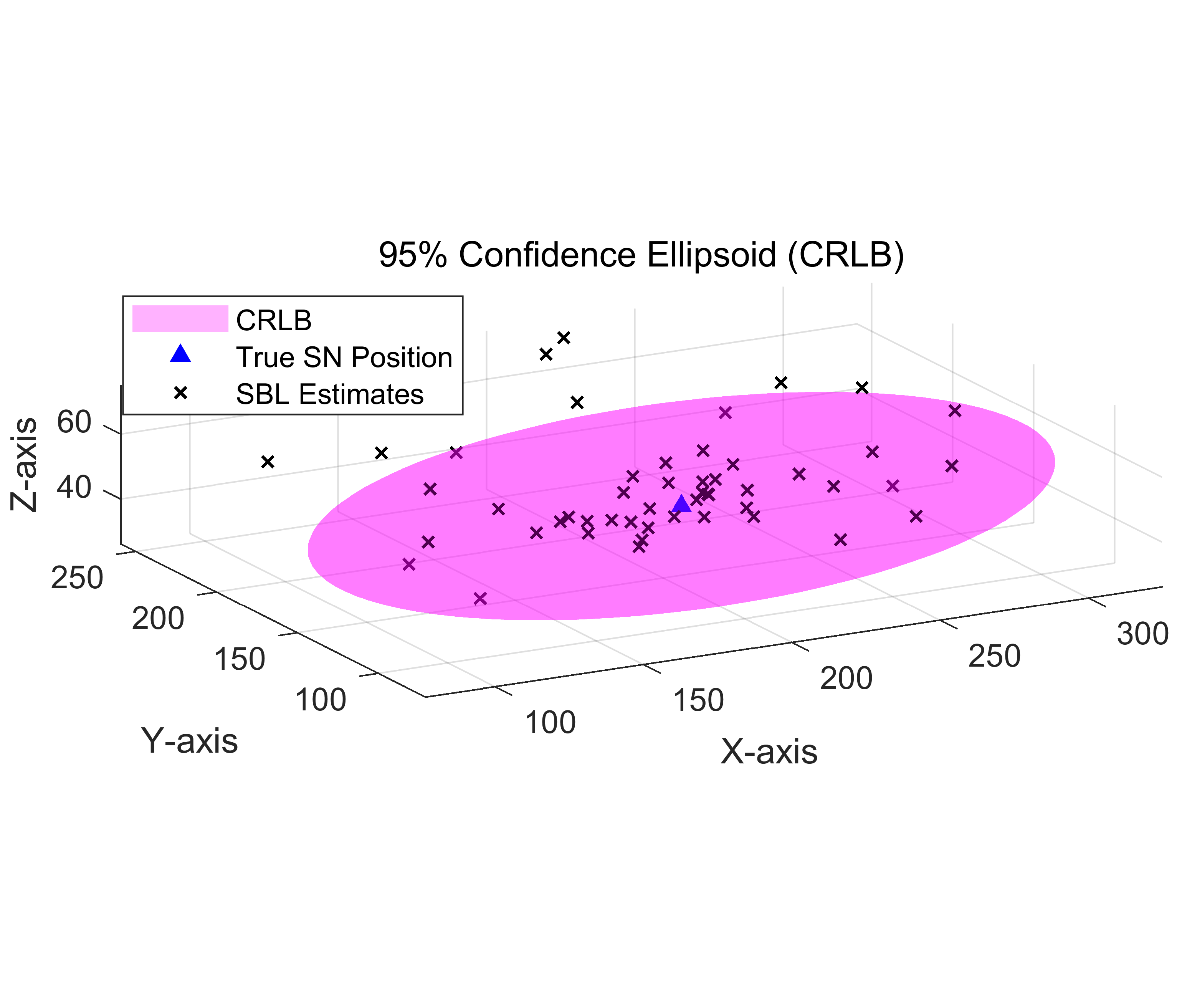}\\
		\end{minipage}%
	}
	\centering
	\caption{The 3-dimensional 95\%-confidence ellipsoid of three schemes based on the CRLB, with 50 SBL estimates  superimposed.}
	\label{fig3}
\end{figure*}
\begin{table}[t]
	\caption{Simulation Parameter Setting.}\label{t1}
	\centering
	\tabcolsep=0.02cm
	\renewcommand\arraystretch{1.3}
	\begin{tabular}{l|c}
		\hline
		\hline
		\textbf{~~~~~~~~~~~~~~~~Parameters} & \textbf{Value} \\
		\hline
		Propagation factor $\epsilon$ \cite{jiang2023underwater} & 1.5  \\
		Acoustic signal frequency $f$& 5kHz \\
		Number of underwater antennas $T$ & 4 \\
		Number of reflection elements $M$ & 512 \\
		Background noise power $\sigma^2$ & -60dBm \\
		Minimum AN power $\sigma_v^2$~~~ & ~~~-60dBm~~~ \\
		Seabed reflection coefficient $\kappa_b$ \cite{weiss2022semi} & 0.85 \\
		Sampling time interval of the ENs $T_s$ & 0.001s \\
		Number of sampling points $N$ & 64 \\
		Speed of underwater sound $c$ & 1500m/s \\
		Seabed depth $h$ & 100m\\
		Position of SN $\mathbf{p}_S$ & (200.7m,140.6m,50.2m)\\
		Position of UARIS $\mathbf{p}_U$ & (500m,210m,30m)\\
		\hline
		\hline
	\end{tabular}
	\label{table1}
\end{table}
\begin{figure}[!t]
	\centering
	\subfigure[Sum-rate of different schemes]{
		\includegraphics[width=0.48\textwidth,height=0.4\textwidth]
		{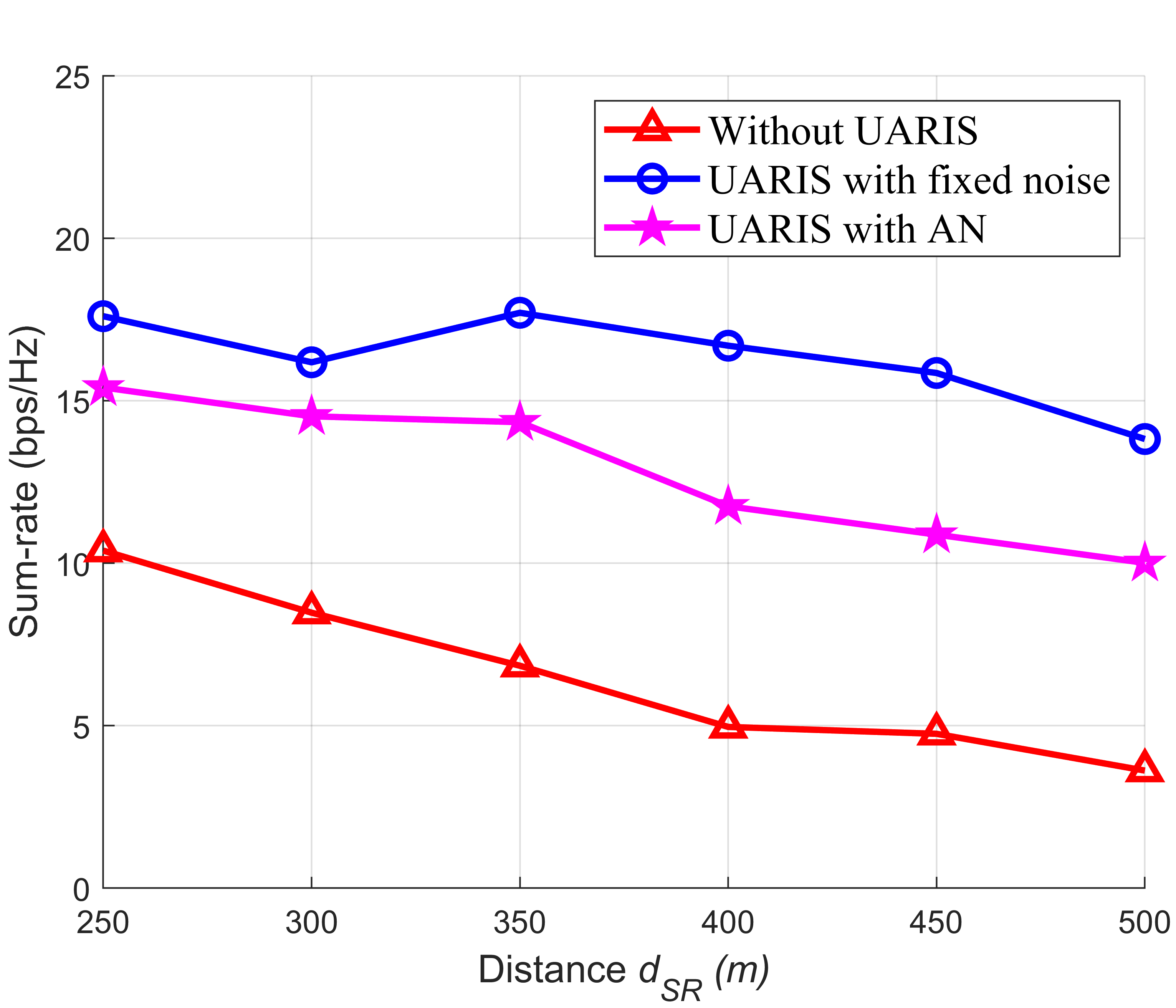}\label{fig3a}
	}
	\centering
	\subfigure[RMS Miss Distance of different schemes]{
		\includegraphics[width=0.48\textwidth,height=0.4\textwidth]{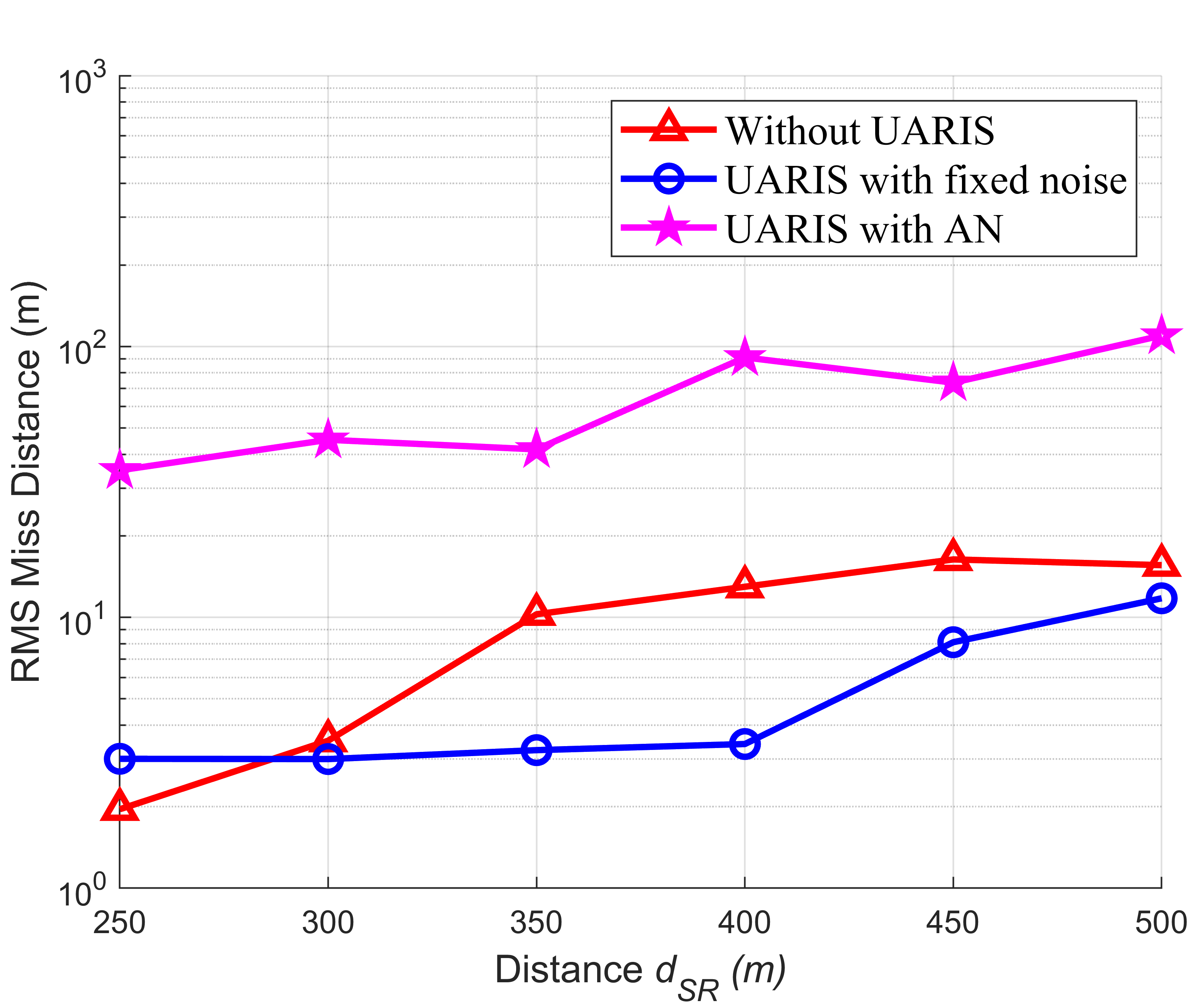} \label{fig3c}
	}
	\DeclareGraphicsExtensions.
	\caption{Coverage performance of different schemes.}
	\label{dif}
\end{figure}
\subsection{Validation of the CRLB}\label{vacrlb}
In this subsection, we validate the CRLB analysis of (\ref{27}). Specifically, the experimental parameters are set as $\xi = 0.02$, $d_{SR} = 300\text{m}$, $d_{ER} = 20\text{m}$ and $P_t^{max} = 30\text{dBm}$. Then, we solve the 95\% confidence ellipses for the three schemes based on (\ref{27}) and overlay the estimated pSBL results from 50 Monte Carlo experiments.

As shown in Fig. \ref{fig3}, due to the CRLB, the empirical results exhibit a good fit with the predicted theoretical accuracy, which aligns with our analytical derivation of the bound. Meanwhile, comparing Fig. \ref{fig4c} with Fig. \ref{fig4a} and Fig. \ref{fig4b}, it can be observed that the proposed scheme significantly reduces the theoretical localization accuracy of the ENs for the SN. This aligns with the inference shown in (\ref{22}), which states that the CRLB of the eavesdropping localization is proportional to the noise introduced by UARIS.
\subsection{Coverage Performance of Different Schemes}\label{cov}

In this subsection, to observe the coverage performance of the proposed UARIS-based new architecture, we plot the relationship between the sum-rate and the average distance $d_{SR}$ between the SN and the RNs by adjusting $d_{SR}$ and conducting Monte Carlo experiments.
	
As shown in Fig. \ref{dif}, compared to the UAC without UARIS, although the case of UARIS with fixed noise significantly improves the sum rate of the underwater communication system, it fails to ensure the location privacy of the SN. When the distance $d_{SR}$ is larger, the Root Mean Square (RMS) miss distance of its localization is smaller than that of the UAC without UARIS. Furthermore, compared to the UAC without UARIS, the proposed scheme not only improves the sum rate of the underwater communication system by 97\%, but also enhances the RMS miss distance of the eavesdropping localization by approximately 5.5 times. Compared to the case of UARIS with fixed noise, although the proposed scheme reduces the sum-rate by 21.4\%, it enhances the RMS miss distance of the eavesdropping localization by approximately 11.2 times. 

These results indicate that the proposed scheme can enhance the communication performance of the underwater communication system while protecting the location privacy of the SN.

\subsection{Effect of the Optimization Weight $\xi$}\label{eff}
As a critical parameter in optimization problem (\ref{p3}), the optimization weight $\xi$ plays a decisive role in the final optimization result. Therefore, to explore the impact of different optimization weights on performance, under the experimental settings of $d_{SR} = 300\text{m}$, $d_{ER} = 20\text{m}$ and $P_t^{max} = 30\text{dBm}$, we conducted Monte Carlo experiments with various values of the weight $\xi$.

\begin{figure}[!t]
	\centering
	\includegraphics[width=3.2in]{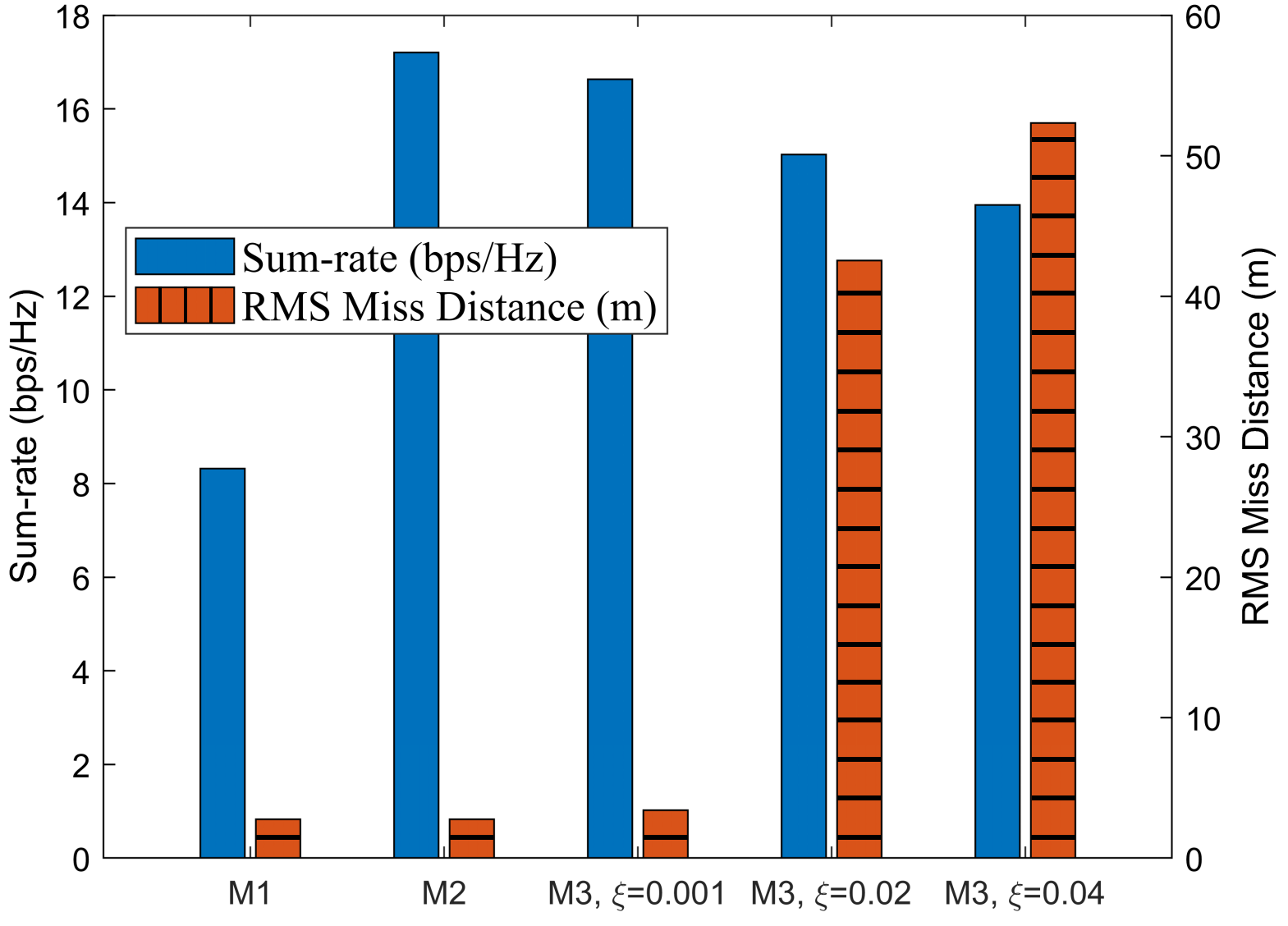}
	\caption{Optimization Performance of the proposed scheme with different weights $\xi$.}
	\label{lam}
\end{figure}

As shown in Fig. \ref{lam}, when the optimization weight is minimal, such as $\xi = 0.001$, the optimization result of the proposed scheme is almost identical to that of UARIS with fixed noise. This indicates that a minimal weight $\xi$ leads the proposed scheme to focus solely on optimizing the sum rate. However, when the optimization weight is larger, such as $\xi = 0.02$ and $\xi = 0.04$, the proposed scheme can improve the sum rate of the UAC system while protecting the SN location privacy. Specifically, when the optimization weight $\xi=0.02$, compared to the case of UARIS with fixed noise, the proposed scheme improves the eavesdropping localization error by approximately 14.5 times while only reducing the sum-rate by 12.7\%. When the optimization weight $\xi=0.04$, compared to Scheme 2, the proposed scheme improves the eavesdropping localization error by approximately 18.0 times while only reducing the communication rate by 18.9\%.

 Additionally, it can be observed from Fig. \ref{lam} that as the weight $\xi$ increases, the optimized system sum rate decreases, while the RMS miss distance of eavesdropping localization increases. This demonstrates that the optimization weight $\xi$ influences the tendency of the proposed scheme toward optimizing the two objectives.
\subsection{Iampact of the Total Power Limit}\label{effe}
\begin{figure}[!t]
	\centering
	\subfigure[Sum-rate of different schemes]{
		\includegraphics[width=0.48\textwidth,height=0.42\textwidth]
		{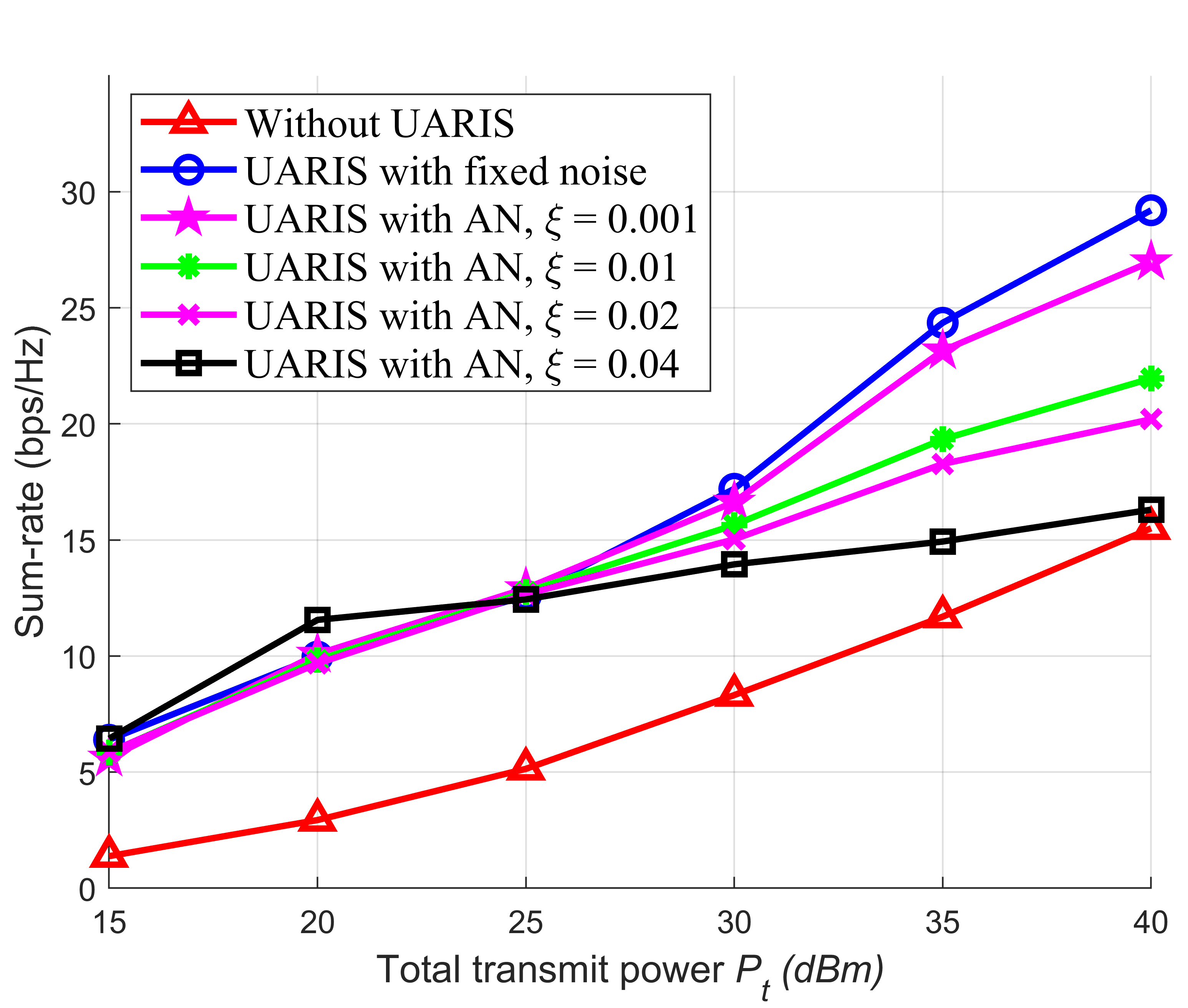}\label{fig5a}
	}
	\centering
	\subfigure[RMS Miss Distance of different schemes]{
		\includegraphics[width=0.48\textwidth,height=0.42\textwidth]{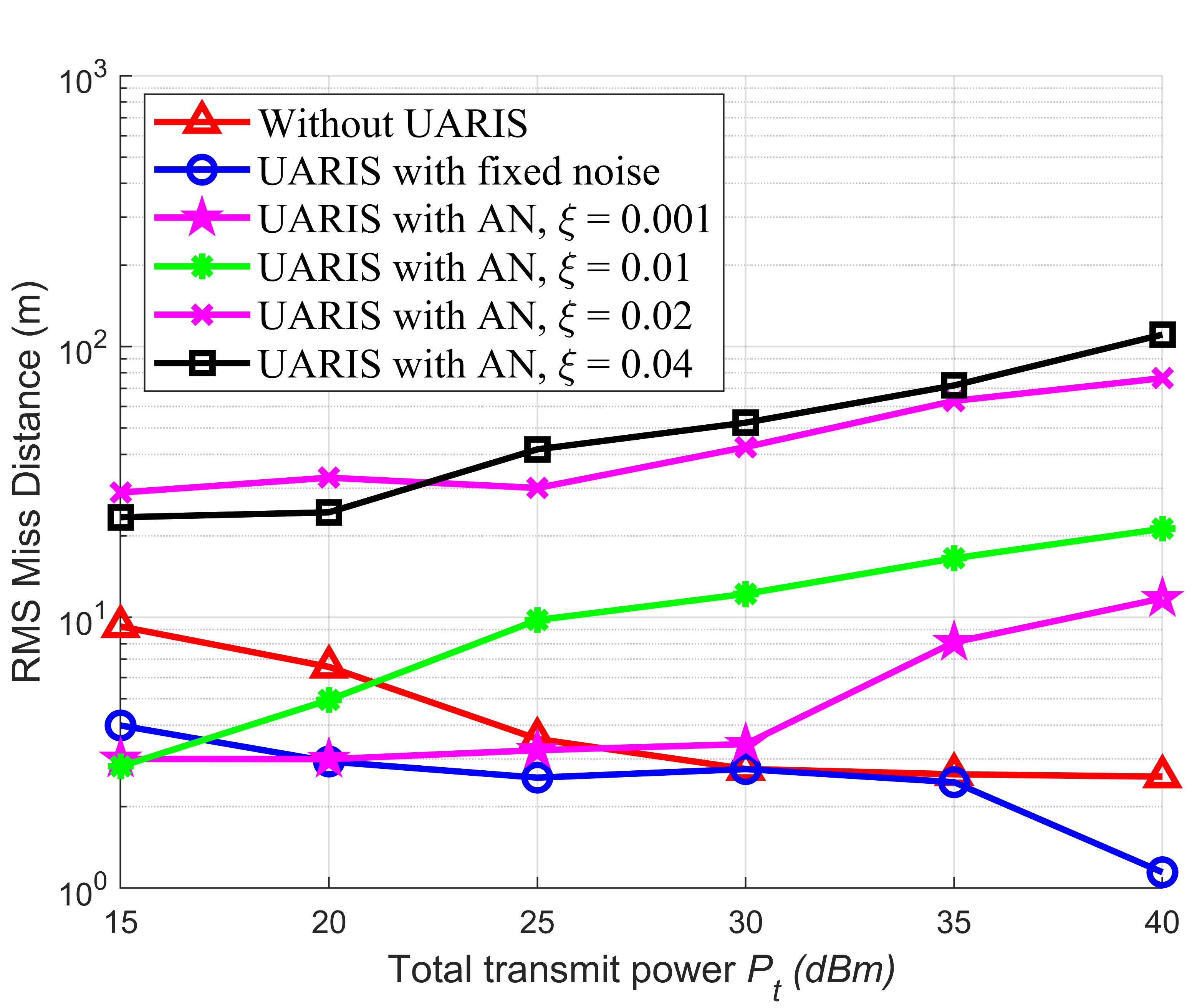} \label{fig5c}
	}
	\DeclareGraphicsExtensions.
	\caption{Optimization performance under different total power limits $P_t^{max}$.}
	\label{P}
\end{figure}

In this subsection, we will evaluate the impact of power levels on the optimization performance of the proposed scheme. Specifically, under the experimental settings of $d_{SR} = 300\text{m}$, $d_{ER} = 20\text{m}$, we adjust the total power limit and conduct Monte Carlo experiments to plot the relationship between the total power limit and the optimization performance.

As shown in Fig. \ref{P}, with the increase in the total power limit $P_t^{max}$, both the UAC without UARIS and the the case of UARIS with fixed noise exhibit an upward trend in system sum rate, while the RMS miss distance of eavesdropping localization decreases. This is because the increase in total system power improves the SNR at the EN, thereby enhancing its localization accuracy. In contrast, the proposed scheme not only increases the sum rate as the total power increases, but also enhances the RMS miss distance of eavesdropping localization. This is because, as the total power increases, UARIS can allocate more power to emit AN. Meanwhile, through UARIS beamforming, it reduces the interference of the added noise on the RN while enhancing the interference on eavesdropping localization. 

Furthermore, as observed in Fig. \ref{P}, when the optimization weight $\xi$ is minimal, specifically $\xi = 0.001$, the proposed scheme is almost unable to protect the location privacy of the SN. On the other hand, when the optimization weight $\xi$ is huge, such as $\xi = 0.04$, the proposed scheme struggles to improve the sum rate. Therefore, The simulation results in Fig. (a) indicate a clear trade-off between communication rate and localization protection. A smaller value of $\xi$ can provide a certain degree of localization protection while maintaining a higher communication rate. In contrast, a larger value of $\xi$ significantly increases the eavesdropper localization error but has a more significant impact on the communication rate. Therefore, the weight should be chosen appropriately according to the specific application scenario to achieve a balance between security and communication efficiency.
\subsection{Impact of the Number of UARIS Elements}\label{effe2}
\begin{figure}[!t]
	\centering
	\subfigure[Sum-rate of different schemes]{
		\includegraphics[width=0.48\textwidth,height=0.42\textwidth]
		{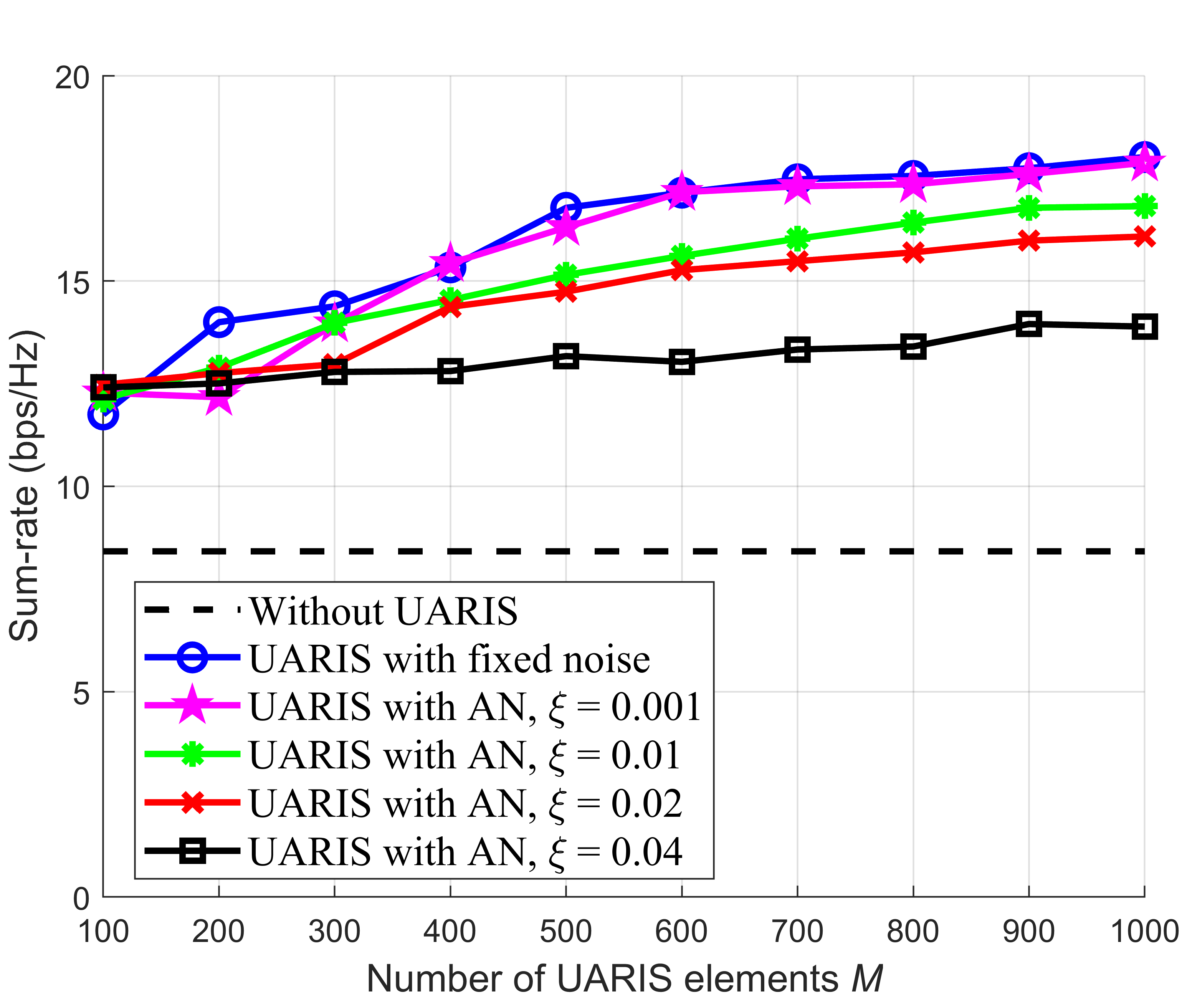}\label{fig6a}
	}
	\centering
	\subfigure[RMS Miss Distance of different schemes]{
		\includegraphics[width=0.48\textwidth,height=0.42\textwidth]{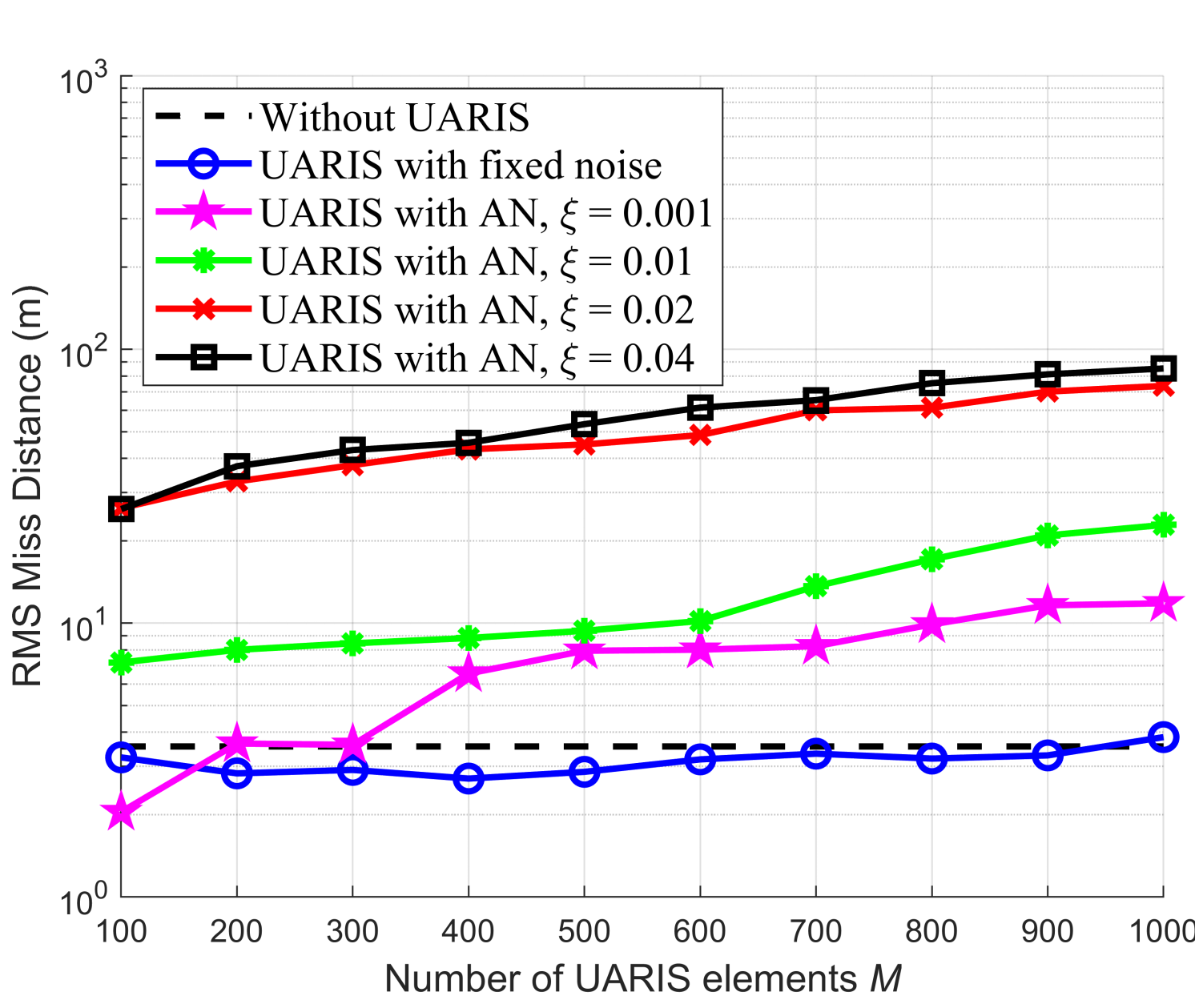} \label{fig6c}
	}
	\DeclareGraphicsExtensions.
	\caption{Performance under different numbers of UARIS elements.}
	\label{RISN}
\end{figure}
In this subsection, we will evaluate the impact of the number of UARIS elements on the optimization performance of the proposed scheme. Specifically, under the experimental settings of $d_{SR} = 300\text{m}$, $d_{ER} = 20\text{m}$ and $P_t^{max} = 30\text{dBm}$, we adjust the number of UARIS elements and conduct Monte Carlo experiments to plot the relationship between the element number and the optimization performance.

As shown in Fig. \ref{fig6a}, with the increase in the number of UARIS reflecting elements $M$, the sum-rate is significantly improved both in the case of UARIS with fixed noise and the proposed scheme with different optimization weight values $\xi$. Furthermore, as observed from Fig. \ref{fig6c}, the number of UARIS elements has no significant impact on the eavesdropping localization error in the case of UARIS with fixed noise. However, the eavesdropping localization error of the proposed scheme increases as the number of UARIS elements increases. This phenomenon is because more UARIS elements can provide finer beamforming and signal control, thereby improving the system communication performance, while saving more energy for transmitting AN to enhance the protection of location privacy.
\section{Conclusion}
\label{concl}
In this paper, we proposed a novel UARIS architecture integrated with an AN module to address the dual challenge of communication enhancement and location privacy protection in shallow water environments. Our system leverages the intelligent control of reflected signals to significantly improve the communication quality while introducing controllable noise to disrupt the attempts to locate the SN. Through the derivation of the CRLB and the formulation of a multi-objective optimization problem, we designed an efficient algorithm based on FP to optimize transmission beamforming, reflective precoding, and noise factor. Simulation results validated that our proposed scheme not only enhances data transmission rates but also effectively safeguards the location privacy of the SN. Specifically, compared to the traditional UARIS, the proposed scheme improved the eavesdropping localization error by approximately 14.5 times while only reducing the sumrate by 12.7\%.

This work conducts an initial exploration into secure and efficient underwater communication systems, particularly in scenarios where privacy concerns are critical. Further research may explore the adaptability of our approach under varying underwater conditions and broader network configurations.
\bibliography{reference}
\end{document}